\documentclass[10pt,a4paper]{article}
\usepackage{float}
\usepackage{graphicx}
\usepackage{amsmath}
\usepackage{amssymb}
\usepackage{latexsym}
\usepackage{color}

\usepackage{caption}
\usepackage{subcaption}
\usepackage{refstyle}
\usepackage{lipsum}

\providecommand{\openone}{\leavevmode\hbox{\small1\kern-4.3pt\normalsize1}}

\begin{document}

\bigskip \thispagestyle{empty}

\begin{center}
\vspace{1.8cm}

\textbf{\large A simple analytical expression of quantum Fisher and Skew information and their dynamics under decoherence channels}


\vspace{1.5cm}

{\bf N. Abouelkhir
}$^{a}$, {\bf H. EL Hadfi}$^{a}$, {\bf A. Slaoui }$^{a,b}${\footnote {email: {\sf abdallah.slaoui@um5s.net.ma}}} and {\bf R. Ahl Laamara}$^{a,b}$

\vspace{0.5cm}

$^{a}${\it LPHE-Modeling and Simulation, Faculty  of Sciences,
	Mohammed V University in Rabat, Rabat, Morocco.}\\
$^{b}${\it Centre of Physics and Mathematics, CPM, Faculty  of Sciences, Mohammed V University in Rabat, Morocco.}\\[1em]

\vspace{1.60cm} \textbf{Abstract}
\end{center}

\baselineskip=18pt \medskip
In statistical estimation theory, it has been shown previously that the Wigner-Yanase skew information is bounded by the quantum Fisher information associated with the phase parameter. Besides, the quantum Cramér-Rao inequality is expressed in terms of skew information. Since these two fundamental quantities are based on the concept of quantum uncertainty, we derive here their analytical formulas for arbitrary two qubit $X$-states using the same analytical procedures. A comparison of these two informational quantifiers for two quasi-Werner states composed of two bipartite superposed coherent states is examined. Moreover, we investigated the decoherence effects on such quantities generated by the phase damping, depolarization and amplitude damping channels. We showed that decoherence strongly influences the quantum criteria during the evolution and these quantities exhibit similar dynamic behaviors. This current work is characterized by the fact that these two concepts play the same role and capture similar properties in quantum estimation protocols.

\vspace{0.25cm}


\section{Introduction}
Quantum correlations other than entanglement are considered a resource in various tasks of quantum information tasks, such as quantum teleportation \cite{Braunstein1998,Bouwmeester1997}, quantum key distribution \cite{Bennett1984}, quantum metrology \cite{Giovannetti2006,Giovannetti2011,Huelga1997} and quantum computing \cite{Datta2011}. For instance, one can significate the role of quantum correlation, or more precisely, quantum entanglement in the field of quantum metrology \cite{Augusiak2016,Hadfi2017}. Indeed, quantum metrology has been going through great development in recent years. This field focuses on making high precision measurements of given parameters using quantum systems and quantum resources \cite{Huelga1997}. In particular, the quantum states with particles exhibiting high quantum entanglement provide a higher precision of physical quantities \cite{Meyer2001,Leibfried2004}. The precision of the measurements refers to existing of two possible scales;  the standard quantum limit given by $\delta\sim N^{-\frac{1}{2}}$ obtained by the measurement with uncorrelated particles and Heisenberg limit rendered by $\delta\sim N^{-1}$, which can be reached by various forms of quantum features including entanglement, with $N$ called the number of quantum resource. In the quantum context from various perspectives, the Heisenberg limit is significantly better than the standard quantum limit. It has been found that the use of $N$ correlated particles can provide high precision measurements in parameter estimation \cite{Huang2016,Riedel2010} as well as the use of the multi-qubit state as probe states outperforms the single-qubit state in parameter estimation \cite{Hu22020}. Various works have been proposed for connecting quantum metrology to quantum information science, in particular to characterize nonclassical correlations \cite{Kim2018,Slaoui2019,SlaouiD2018} and its interconnections with quantum coherence measures have also been identified \cite{Hu2018}.\par

Quantum metrology is a discipline that combines quantum mechanics and classical metrology \cite{Paris2009}. This emerging field of physics focuses on the theoretical and experimental study of metrology protocols \cite{Toth2014}. The main objective is to determine the precision with which the true value of an unknown parameter can be estimated. In particular, this field studies how particular properties of quantum mechanics affect the accuracy in the estimation of one or more parameters \cite{Szczykulska2016}. The fundamental theoretical tools for these studies are provided by quantum estimation theory, which gives fundamental bounds for the accuracy attainable in terms of quantum Fisher information. In this frame, quantum Fisher information is employed as the tool to achieve the desired accuracy in quantum estimation protocols. In fact, it has turned out that quantum Fisher information and skew information in the physical system has practical significance in quantum metrology \cite{Luo2003,Luo2004}. In a quantum metrological process, the determination of some unknown parameter can be achieved by measuring a probe system whose quantum state depends on that parameter. This process contains four steps; Firsy, the preparation of the initial state. Second, parametrization by unitary evolution. Third, the quantum measurement on the final state and finally the classical estimation which can be achieved by measuring an appropriate observable in the output state. Outside of quantum metrology, the quantum Fisher information also connects to other aspects of quantum physics, such as quantum phase transition \cite{Ye2016,ACarollo2020}, information geometry \cite{Girolami2016}, entanglement witness \cite{Hauke2016} and quantum thermodynamics \cite{Hasegawa2020}. On the other hand, skew information \cite{Wigner1963} characterizes the information content of quantum systems and also plays an important role in quantum estimation, in particular, it is applied to improve the accuracy of interferometric phase estimation \cite{Girolami2013}. This measure was introduced by Wigner and Yanase to quantify the non-commutativity between the quantum state $\rho$ and a self-adjoint operator $K$. Furthermore, it can be interpreted as a measure of quantum uncertainty \cite{Luo2004} and used to quantify non-classical correlations for bipartite systems \cite{Luo2012,Slaoui2018}.\par

Frequently, the quasi-Werner states are used in quantum correlation \cite{Mishra2016,Chatterjee2021} and coherence \cite{Ali2021} studies. A comparative study of quantum discord and quantum entanglement for quasi-Werner states to capture the sensitivity in quantum metrology is studied in Ref.\cite{Khalid2018}. These states can be considered as a statistical mixture of a maximally entangled pure state and a maximally mixed state. Two of these four bipartite entangled coherent states are maximally entangled and they form perfect Werner states and the other two states (non-maximally entangled states) are defined as quasi-Werner states. Some of these states have been used recently to develop quantum technology such as in quantum key distribution \cite{Srikara2020} and quantum metrology \cite{Malpani2019}. This motivated us to use quasi-Werner states to address the pertinence and efficiency of skew information to detect the precision of the estimated parameters and to compare it with quantum Fisher information.\par 

On the other hand, when the physical systems concerned are not in themselves particularly complex but interact with external degrees of freedom beyond our control, where these interactions are least likely to be negligible, the global dynamics of these systems are difficult to take into account quantitatively \cite{Mazzola2010,Tegmark2000}. In these situations, physical systems are invariably subject to noise which leads to the degradation of quantum phenomena. Due to this practical limitation, the dynamics of quantum correlations have been investigated from different aspects and significant progress has been achieved to protect them while passing through quantum channels \cite{Hu2020,Hu2016,Hu2015}. The loss of these phenomena under the influence of the external environment is known as decoherence which has implications for quantum information processing tasks \cite{Burkov2007,Seo2016}. This problem is the subject of the theory of open quantum systems, a rich and rapidly expanding current research field. Briefly, the interaction between a quantum system and its environment can result in the dissipation or loss of information contained in the system to its environment. Depending on the type of interaction, the dynamics can be Markovian \cite{Breuer2002,Spohn1980} where the coupling strength of the system with the environment is weak, or non-Markovian in which the evolution process exhibits the behavior of memory effects \cite{Breuer2016,Vega2017}. Here, we perform a dynamic investigation of quantum Fisher information and skew information under various Markovian channels operated on the quasi-Werner states based on bipartite superposed coherent states ($\sim|\alpha,\beta\rangle\pm|-\alpha,-\beta\rangle$). Analytical results for these two information quantities under the phase damping channel, the depolarizing channel and the amplitude damping channel are derived.\par 

This paper is organized as follows. In section $2$, we give some basic properties of two-qubit $X$-states and how to calculate quantum Fisher information and skew information for them. In Section $3$, we provide the explicit expressions for these two quantities under three different decoherence channels; phase damping channel, depolarizing channel and amplitude damping channel. The dynamics of quantum Fisher information and skew information for two quasi-Werner states evolving in a noisy environment are also studied and numerical results are given. To examine the role of quantum entanglement in improving the precision of quantum metrology protocols, we quantify the entanglement amount of two quasi-Werner states using Wootters concurrence. Finally, we close this present investigation in the last section.
\section{Analytical expression of Fisher and skew information for two qubit $X$-states}
\subsection{Two qubits $X$-state density matrix }
Actually, two-qubit $X$-states have the virtue of being able to perform many calculations analytically, which is always an aid to understanding and application. Moreover, they have already found powerful applications in many analyses concerning entanglement \cite{Yu2004,Sun2017} and quantum discord \cite{Fanchini2010,Maziero2010}. Their visual appearance, which resembles the letter $X$ with non-zero density matrix elements only along the diagonal and anti-diagonal, has led to them being called $X$-states. Thereby, these include a variety of states of interest that are classical or non-classical, separable or non-separable, such that limiting our investigation to them is not too restrictive. The extension to multi-qubit states encompasses many interesting states in quantum information processing such as Greenberger-Horne-Zeilinger \cite{Greenberger1990} and Werner \cite{Dur2000} states for three-qubit states and Dicke states \cite{Guhne2009} for $N$-qubits. Here, we will focus mainly on two-qubit states $\rho_{AB}$ whose density matrices are $X$-shaped and having the following form
\begin{equation}\label{9}
	\begin{aligned}
		\rho_{AB}=& \frac{1}{4}\left(\mathcal{T}_{00} \sigma_{0} \otimes \sigma_{0}+\mathcal{T}_{33} \sigma_{3} \otimes \sigma_{3}+\mathcal{T}_{30} \sigma_{3} \otimes \sigma_{0}+\mathcal{T}_{03} \sigma_{0} \otimes \sigma_{3}\right.\\
		&\left.+\mathcal{T}_{11} \sigma_{1} \otimes \sigma_{1}+\mathcal{T}_{22} \sigma_{2} \otimes \sigma_{2}+\mathcal{T}_{12} \sigma_{1} \otimes \sigma_{2}+\mathcal{T}_{21} \sigma_{2} \otimes \sigma_{1}\right),
	\end{aligned}
\end{equation}
in Fano-Bloch representation, where $\mathcal{T}_{\alpha \beta}=\operatorname{Tr}\left(\rho_{AB}\sigma_{\alpha} \otimes \sigma_{\beta}\right)$ are the correlation matrix elements, with $\sum_{i=1}^{4}\rho_{ii}=1$, $\rho_{22}\rho_{33}\geq\mid\rho_{23}\mid^{2}$ and $\rho_{11}\rho_{44}\geq\mid\rho_{14}\mid^{2}$ are fulfilled. This density matrix can be written also as
\begin{equation}\label{D}
	\rho_{AB}=\rho_{AB}^{(1)}\oplus\rho_{AB}^{(2)},
\end{equation}
where the sub-matrices $\rho_{AB}^{(1)}$ and $\rho_{AB}^{(2)}$ are defined as
\begin{equation}\label{12}
	\rho_{AB}^{(1)}=\left(\begin{array}{ll}
		\rho_{11} & \rho_{14} \\
		\rho_{41} & \rho_{44}
	\end{array}\right), \hspace{2cm} \rho_{AB}^{(2)}=\left(\begin{array}{ll}
		\rho_{22} & \rho_{23} \\
		\rho_{32} & \rho_{33}
	\end{array}\right),
\end{equation}
in the basis $\mathcal{B}_{1}=\{|00\rangle,|11\rangle\}$ and $\mathcal{B}_{2}=\{|01\rangle,|10\rangle\}$, respectively. We then introduce the generators $\vartheta_{i}$ and $\tilde{\vartheta}_{i}$ ($i=0,1,2,3$) as
\begin{equation}\label{13}
	\begin{aligned}
		&\vartheta_{0}=\left|00\right\rangle \left\langle 00\right| +\left|11\right\rangle \left\langle 11\right|, \hspace{1cm} \tilde{\vartheta}_{0}=\left| 01\right\rangle \left\langle 01\right|+\left| 10\right\rangle \left\langle 10\right|, \\
		&\vartheta_{1}=\left|00\right\rangle \left\langle11\right|+\left| 11\right\rangle \left\langle 00\right|, \hspace{1cm} \tilde{\vartheta}_{1}=\left| 01\right\rangle \left\langle 10\right|+\left|10\right\rangle \left\langle 01\right|, \\
		&\vartheta_{2}=i\left|11\right\rangle \left\langle 00\right|-i\left|00\right\rangle \left\langle11\right|, \hspace{1cm}\tilde{\vartheta}_{2}=i\left| 10\right\rangle \left\langle 01\right|-i\left|01\right\rangle \left\langle 10\right|,\\
		&\vartheta_{3}=\left|00\right\rangle \left\langle00\right|-\left| 11\right\rangle \left\langle 11\right|, \hspace{1cm} \tilde{\vartheta}_{3}=\left| 01\right\rangle \left\langle 01\right|-\left|10\right\rangle \left\langle 10\right|,
	\end{aligned}
\end{equation}
to simplify the derivation of quantum Fisher and skew information. It becomes clear that the matrices $\rho_{AB}^{(1)}$ and $\rho_{AB}^{(2)}$, in this decompostion, takes the form
\begin{equation}\label{15}
	\rho_{AB}^{(1)}=\frac{1}{2} \sum_{\alpha=0}^{3} \chi_{\alpha} \vartheta_{\alpha}, \hspace{2cm} \rho_{AB}^{(2)}=\frac{1}{2} \sum_{\alpha=0}^{3} \tilde{\chi}_{\alpha} \tilde{\vartheta}_{\alpha},
\end{equation}
where ${\chi}_{\alpha}$ and $\tilde{\chi}_{\alpha}$ are given respectively by
\begin{equation}\label{16}
		\chi_{0}=\frac{1}{2}\left(\mathcal{T}_{00}+\mathcal{T}_{33}\right), \hspace{0.75cm} \chi_{1}=\frac{1}{2}\left(\mathcal{T}_{11}-\mathcal{T}_{22}\right), \hspace{0.75cm}
		\chi_{2}=\frac{1}{2}\left(\mathcal{T}_{12}+\mathcal{T}_{21}\right),\hspace{0.75cm} \chi_{3}=\frac{1}{2}\left(\mathcal{T}_{30}+\mathcal{T}_{03}\right),
\end{equation}
and
\begin{equation}\label{17}
		\tilde{\chi}_{0}=\frac{1}{2}\left(\mathcal{T}_{00}-\mathcal{T}_{33}\right), \hspace{0.75cm} \tilde{\chi}_{1}=\frac{1}{2}\left(\mathcal{T}_{11}+\mathcal{T}_{22}\right), \hspace{0.75cm} \tilde{\chi}_{2}=\frac{1}{2}\left(\mathcal{T}_{21}-\mathcal{T}_{12}\right), \hspace{0.75cm} \tilde{\chi}_{3}=\frac{1}{2}\left(\mathcal{T}_{30}-\mathcal{T}_{03}\right).
\end{equation}
\subsection{Explicit form of quantum Fisher information}
Let us consider a quantum state $\rho_{\theta}$ that depends on an unknown parameter $\theta$. If we intend to extract information about $\theta$ from the density matrix $\rho_{\theta}$, we need to perform a set of quantum measurements $\left\lbrace\Pi_{x}\right\rbrace$. In classical metrology \cite{Giovannetti2006,Paris2009}, the Fisher information provides the lower limit of the variance of an unbiased estimator for the estimated parameter. It is defined by
\begin{equation}
F_{c}\left(\theta\right) =\int p\left(x/\theta\right)\left[\frac{\partial\ln p\left(x/\theta\right)}{\partial\theta}\right]^{2} dx, \label{CF}
\end{equation}
where $p\left(x/\theta\right)={\rm Tr}\left[\Pi_{x}\rho_{\theta} \right]$ represents the conditional probability of obtaining the measurement outcome $x$ when the true value of the parameter is $\theta$. According to classical statistical theory, the ultimate accuracy ${\rm Var}\left(\theta \right)$ in estimating $\theta$ from $n$ measurements is restricted by the Fisher information $F_{c}\left(\theta\right)$ via the Cramér-Rao inequality as
\begin{equation}
{\rm Var}\left(\theta\right)\geq\frac{1}{nF_{c}\left(\theta\right)}.
\end{equation}
The quantum Fisher information is defined as the optimization of the classical Fisher information (Eq.(\ref{CF})) over all possible measurements $\left\lbrace\Pi_{x}\right\rbrace$, i.e.,
\begin{equation}
\mathcal{F}\left(\rho_{\theta}\right)=\max_{\Pi_{x}}F_{c}\left(\theta\right).
\end{equation}
Then we can rewrite the above equation explicitly as
\begin{equation}
\mathcal{F}\left(\rho_{\theta}\right)=\operatorname{Tr}(\rho_{\theta} \mathcal{L}_{\theta}^{2})=\operatorname{Tr}\left[\left(\partial_{\theta} \rho_{\theta}\right) \mathcal{L}_{\theta}\right],
\end{equation}
where  $\mathcal{L}_{\theta}$ is so-called symmetric logarithmic
derivative operator and is determined explicitly by the relation
\begin{equation}\label{23}
	\frac{\partial\rho_{\theta}}{\partial\theta}=\frac{1}{2}\left[\rho_{\theta}\mathcal{L}_{\theta}+\mathcal{L}_{\theta}\rho_{\theta}\right].
\end{equation}
Indeed, the quantum Fisher information represents a central role in quantum metrology. It can be used to extract information about some parameters that cannot be measured directly \cite{Paris2009}. Specifically, this quantity is proposed to measure the phase sensitivity of quantum systems. \cite{Yu2018,Yi2012,Liu2016}. It is necessary to mention that this quantity can be derived directly when the symmetric logarithmic derivative operator is explicitly calculated. Here, we restrict our description to the explicit expression of the quantum Fisher information for two-qubit $X$ states (of the form (\ref{9})). The solution of the equation (\ref{23}) leads to the analytical formula of the symmetric logarithmic derivative operator in terms of exponentiation of the density matrix as
\begin{equation}\label{24}
	\mathcal{L}_{\theta}=2 \int_{0}^{\infty} e^{-\rho_{\theta} s}(\partial_{\theta} \rho_{\theta}) e^{-\rho_{\theta} s} ds.
\end{equation}
Using the decomposition $\rho_{\theta}^{n}=\rho_{\theta}^{(1)n}\oplus{\rho_{\theta}^{(2)n}}$ (with $n \geq 1$) and $e^{-\rho_{\theta} s}=\sum_{n=0}^{\infty}\frac{\left(-s \right)^{n}}{n!}\rho_{\theta}^{n}=e^{-\rho_{\theta}^{(1) }s}+e^{-\rho_{\theta}^{(2)}s}$, the expression of the operator $\mathcal{L}_{\theta}$ can then be written as
\begin{equation}\label{25}
	\mathcal{L}_{\theta}=L_{\theta}^{(1)}+ L_{\theta}^{(2)},
\end{equation}
where
\begin{equation}\label{26}
	L_{\theta}^{(1)}= p^{0} \vartheta_{0}+\sum_{i=1}^{3} p^{i} \vartheta_{i}, \hspace{1cm}{\rm and}\hspace{1cm} L_{\theta}^{(2)}=\tilde{p}^{0} \tilde{\vartheta}_{0}+\sum_{i=1}^{3} \tilde{p}^{i} \tilde{\vartheta}_{i}.
\end{equation}
In this respect, the quantum Fisher information can be decomposed as $\mathcal{F}\left(\rho_{\theta}\right)=\mathcal{F}\left(\rho_{\theta}^{(1)}\right)+\mathcal{F}\left(\rho_{\theta}^{(2)}\right)$, where $\mathcal{F}\left(\rho_{\theta}^{(1)}\right)$ (or $\mathcal{F}\left(\rho_{\theta}^{(2)}\right)$) denotes the quantum Fisher information for $\rho_{\theta}^{(1)}$ (or $\rho_{\theta}^{(2)}$) which is expressed as
\begin{equation}\label{28}
	\mathcal{F}\left(\rho_{\theta}^{(1)}\right)=\operatorname{Tr}\left((\partial_{\theta} \rho_{\theta}^{(1)}) L_{\theta}^{(1)}\right), \hspace{1cm}{\rm and}\hspace{1cm} \mathcal{F}\left(\rho_{\theta}^{(2)}\right)=\operatorname{Tr}\left((\partial_{\theta} \rho_{\theta}^{(2)}) L_{\theta}^{(2)}\right).
\end{equation}
Based on the expression of the symmetric logarithmic derivative operator (\ref{26}) for the state $\rho_{\theta}^{(1)}$ and Eq.(\ref{15}), one check that
\begin{equation}\label{29}
	\mathcal{F}\left(\rho_{\theta}^{(1)}\right)=p_{0}\left(\partial_{\theta} \chi_{0}\right)+\sum_{i=0}^{3}\left(p_{i} \partial_{\theta} \chi_{i}\right),
\end{equation}
with $p_{0}$ and $p_{i}$ are completely determined by the following relation
\begin{equation}\label{30}
	\frac{\partial\rho_{\theta}^{(1)}}{\partial\theta}=\frac{1}{2}\left(\rho_{\theta}^{(1)} L_{\theta}^{(1)}+L_{\theta}^{(1)}\rho_{\theta}^{(1)}\right),
\end{equation}
By substituting the expression of the density matrix $\rho_{\theta}^{(1)}$ (eq.\ref{15}) in the expression of the operator $L_{\theta}^{(1)}$ (eq.\ref{26}), we can easily show that
\begin{equation}\label{31}
	\partial_{\theta} \chi_{0}=\chi_{0} p_{0}+\sum_{i=1}^{3} \chi_{i} p_{i}, \hspace{1cm}{\rm and}\hspace{1cm} \partial_{\theta} \chi_{i}=p_{0} \chi_{i}+\chi_{0} p_{i},
\end{equation}
with $i=0,1,2,3$. By solving the above equations, one gets
\begin{equation}\label{32}
		p_{0}=\frac{\chi_{0} \partial_{\theta} \chi_{0}-\sum_{i=1}^{3} \chi_{i} \partial_{\theta} \chi_{i}}{\chi_{0}^{2}-\sum_{i=1}^{3} \chi_{i}^{2}},\hspace{1cm}{\rm and}\hspace{1cm} p_{i}=\frac{1}{\chi_{0}}\left[\frac{g_{\alpha \beta} \chi^{\alpha}\left(\chi^{\beta} \partial_{\theta} \chi^{i}-\chi^{i} \partial_{\theta} \chi^{\beta}\right)}{g_{\alpha \beta} \chi^{\alpha} \chi^{\beta}}\right],
\end{equation}
where $g_{\alpha \beta}={\rm diag}\left(1,-1,-1,-1\right)$ denotes the metric of the Minkowski space-time. Substituting the equations (\ref{32}) and (\ref{31}) into Eq.(\ref{29}), $\mathcal{F}(\rho_{\theta}^{(1)})$ can be expressed as
\begin{equation}\label{34}
	\mathcal{F}\left(\rho_{\theta}^{(1)}\right)=\frac{\left(\partial_{\theta} \chi_{0}\right)^{2}}{\chi_{0}}+\frac{1}{\chi_{0}}\left[\frac{\left(g_{\alpha \beta} \chi^{\alpha} \partial_{\theta}
		\chi^{\beta}\right)^{2}}{g_{\alpha \beta} \chi^{\alpha} \chi^{\beta}}-g_{\alpha \beta}\left(\partial_{\theta} \chi^{\alpha}\right)\left(\partial_{\theta}
	\chi^{\beta}\right)\right].
\end{equation}
The above equation (\ref{34}) is applicable for mixed states (with $\chi_{0}\neq0$ and $g_{\alpha\beta}\chi^{\alpha}\chi^{\beta}\neq0$). For pure states, we have $\rho_{\theta}^{(1)^2}=\rho_{\theta}^{(1)}$ and the symmetric logarithmic derivative operator can be directly calculated as
\begin{equation}\label{L pure}
	L_{\theta}^{(1)}=2\partial_{\theta}\rho_{\theta}^{(1)}.
\end{equation}
Especially, by substituting Eqs (\ref{L pure}) and (\ref{15}) into Eq.(\ref{29}), the quantum Fisher information $\mathcal{F}(\rho_{\theta}^{(1)})$ (\ref{pure1}) reduces to
\begin{equation}\label{pure1}
	\mathcal{F}\left(\rho_{\theta}^{(1)}\right)=\left(\chi_{0}\right)^{2}+\sum_{i=1}^{3}\left(\chi_{i}\right)^{2}.
\end{equation}
Consequently, the quantum Fisher information $\mathcal{F}\left(\rho_{\theta}^{(2)}\right)$ associated with the matrix $\rho_{\theta}^{(2)}$ (\ref{12}) can be determined similarly. It writes as
\begin{equation}\label{35}
	\mathcal{F}\left(\rho_{\theta}^{(2)}\right)=\frac{\left(\partial_{\theta} \tilde{\chi}_{0}\right)^{2}}{\tilde{\chi}_{0}}+\frac{1}{\tilde{\chi}_{0}}\left[\frac{\left(g_{\alpha \beta} \tilde{\chi}^{\alpha} \partial_{\theta} \tilde{\chi}^{\beta}\right)^{2}}{g_{\alpha \beta} \tilde{\chi}^{\alpha}
		\tilde{\chi}^{\beta}}-g_{\alpha \beta}\left(\partial_{\theta} \tilde{\chi}^{\alpha}\right)\left(\partial_{\theta} \tilde{\chi}^{\beta}\right)\right],
\end{equation}
and it can be simplified to
\begin{equation}\label{pure2}
	\mathcal{F}\left(\rho_{\theta}^{(2)}\right)=\left(\tilde{\chi_{0}}\right)^{2}+\sum_{i=1}^{3}\left(\tilde{\chi_{i}}\right)^{2},
\end{equation}
for pure states. unlike calculation methods that rely on the diagonalization of the density matrix, this new expression (\ref{35}) bypasses matrix diagonalization and can be successfully applied in various quantum systems.

\subsection{Explicit form of Wigner-Yanase skew information}
Typically, skew information is widely regarded as a special type of quantum Fisher information since it falls into the category of monotonic metrics on the state space \cite{Petz1996,Petz21996}. Moreover, it possesses many geometric features analogous to those of Fisher information, both based on the notion of quantum uncertainty as well as Cramér-Rao inequality can be rewritten in terms of skew information \cite{Petz2002,Petz31996,Luo22004}. This motivated us to investigate the similarity of these two quantities. Usually, the total uncertainty of a single observable $K$ in a quantum state $\rho_{\theta}$ is quantified by the variance as ${\rm Var}\left(\rho_{\theta},K\right)={\rm Tr}\left(\rho_{\theta}K^{2}\right)-\left({\rm Tr}\left(\rho_{\theta}K\right) \right)^{2}$. Due to the classical ignorance on the classical mixture in $\rho_{\theta}$, the uncertainty quantification in terms of variance includes the classical ignorance ${\rm Var}_{c}\left(\rho_{\theta}\right)$ and a quantum part ${\rm Var}_{q}\left(\rho_{\theta}\right)$ for the mixed states, such that ${\rm Var}\left(\rho_{\theta},K\right)={\rm Var}_{c}\left(\rho_{\theta}\right)+{\rm Var}_{q}\left(\rho_{\theta}\right)$. To deal only with the quantum part of the variance, Wigner and Yanase \cite{Wigner1963} introduced the following quantity
\begin{equation}\label{X}
\mathcal{I}\left(\rho_{\theta}\right)=-4{\rm Tr}[\sqrt{\rho_{\theta}},K]^{2},
\end{equation}
called the Wigner-Yanase skew information, as a measure of the information content on the values of observables not commuting with $K$. In the situation where the quantum estimation is carried out in an interferometric configuration (i.e., the estimation of an unknown phase shift) \cite{SlaouiDrissi}, the density matrix is generated by the unitary dynamics $\rho_{\theta}=e^{-iK\theta}\rho_{AB}e^{iK\theta}$ and $\rho_{\theta}$ obeys the Landau-von Neumann equation $i\partial_{\theta}\rho_{\theta}=\left[K,\theta_{\theta}\right]$. Thus, the above equation (\ref{X}) becomes
\begin{equation}\label{36}
\mathcal{I}\left(\rho_{\theta}\right)=4{\rm Tr}\left(\partial \sqrt{\rho_{\theta}}\right)^{2}.
\end{equation}
This quantity can be regarded as a measure of the non-commutativity degrees between a considered state $\rho_{\theta}$ and the observable $K$ and satisfies all the known criteria for an information-theoretic measure; It reduces to the variance for pure states and vanishes when the density matrix commutes with the conserved quantity $K$. It is invariant as long as the state changes for isolated systems according to the Landau-von Neumann. Another interesting property of skew information is its additive, i.e. the information content of a multiparty system is independent of the sum of the information provided by the subsystems \cite{Liu2016}. For two qubits $X$-state (\ref{D}), the skew information (\ref{36}) can be equivalently represented as
\begin{equation}\label{37}
	\mathcal{I}\left(\rho_{\theta}\right)=\mathcal{I}\left(\rho_{\theta}^{(1)}\right) +\mathcal{I}\left(\rho_{\theta}^{(2)}\right),
\end{equation}
where
\begin{equation}\label{38}
	\mathcal{I}\left(\rho_{\theta}^{(1)}\right)=4 \operatorname{Tr}\left(\partial_{\theta} \sqrt{\rho_{\theta}^{(1)}}\right)^{2}, \hspace{1cm}{\rm and}\hspace{1cm} \mathcal{I}\left(\rho_{\theta}^{(2)}\right)=4 {\rm Tr}\left(\partial_{\theta} \sqrt{\rho_{\theta}^{(2)}}\right)^{2}.
\end{equation}
Clearly, to obtain the explicit form of $\mathcal{I}\left(\rho_{\theta}^{(1)}\right)$ in
terms of the coefficients characterizing the state under consideration, we use the same procedure as the one we adopted to get the quantum Fisher information. First, it easy to see that the matrices $\sqrt{\rho_{\theta}^{(1)}}$ and $\sqrt{\rho_{\theta}^{(2)}}$ can be expanded as
\begin{equation}\label{39}
		\sqrt{\rho_{\theta}^{(1)}}=t_{0} \vartheta_{0}+\sum_{i=1}^{3} t^{i} \vartheta_{i}, \hspace{1cm}{\rm and}\hspace{1cm}
		\sqrt{\rho_{\theta}^{(2)}}=\tilde{t}_{0} \tilde{\vartheta}_{0}+\sum_{i=1}^{3}
		\tilde{t}^{i} \tilde{\vartheta}_{i}.
\end{equation}
Based on above equations, one shows that
\begin{equation}\label{40}
		\mathcal{I}\left(\rho_{\theta}^{(1)}\right)=8\left[\left(\partial_{\theta}
		t_{0}\right)^{2}+\sum_{i=1}^{3}\left(\partial_{\theta} t^{i}\right)^{2}\right],\hspace{1cm}{\rm and}\hspace{1cm}
		\mathcal{I}(\rho_{\theta}^{(2)})=8\left[\left(\partial \tilde{t}_{0}\right)^{2}+
		\sum_{i=1}^{3}\left(\partial \tilde{t}^{i}\right)^{2}\right],
\end{equation}
So, it becomes clear that to calculate $\mathcal{I}\left(\rho_{\theta}^{(1)}\right)$ and $\mathcal{I}(\rho_{\theta}^{(2)})$, we need to evaluate the coefficients
$t_{0}$ ($\tilde{t}_{0}$) and $t_{i}$ ($\tilde{t}_{i}$) in terms of $\chi_{0}$ ($\tilde{\chi}_{0}$) (\ref{16}) and $\chi_{i}$ ($\tilde{\chi}_{i}$) (\ref{17}), respectively. For the matrix $\sqrt{\rho_{\theta}^{(1)}}$, we find
\begin{equation}\label{41}
	\rho_{\theta}^{(1)}=\left(t_{0}^{2}+\left|t^{2}\right|\right) \vartheta_{0}+2 t_{0} \sum_{i=1}^{3} t^{i} \vartheta_{i},
\end{equation}
with $t=(t_{1},t_{2},t_{3})$. By comparing the result (\ref{41}) with the expanded form of the matrix $\rho_{AB}^{(1)}$ (\ref{15}), on obtains
\begin{equation}\label{42}
	t_{0}^{2}+|t|^{2}=\frac{\chi_{0}}{2}, \hspace{2cm}2 t_{0} t^{i}=\frac{\chi_{i}}{2},
\end{equation}
and the solutions $t_{0}$ and $t_{i}$ reads as
\begin{equation}\label{43}
	t_{0}=\frac{1}{2} \sqrt{\chi_{0}+\sqrt{\chi_{0}^{2}-|\chi|^{2}}}, \hspace{1cm}{\rm and}\hspace{1cm} t^{i}=\frac{1}{2} \frac{\chi_{i}}{\sqrt{\chi_{0}+\sqrt{\chi_{0}^{2}-|\chi|^{2}}}},
\end{equation}
with $\chi=(\chi_{1}, \chi_{2}, \chi_{3})$. We end up with
\begin{equation}\label{44}
		\partial_{\theta} t_{0}=\frac{1}{4\sqrt{\chi_{0}^{2}-|\chi|^{2}}}\left[\sqrt{\chi_{0}+\sqrt{\chi_{0}^{2}-|\chi|^{2}}}\left(\partial_{\theta} \chi_{0}\right)-\frac{\Sigma_{j=1}^{3} \chi_{j}\left(\partial_{\theta} \chi_{j}\right)}{\sqrt{\chi_{0}+\sqrt{\chi_{0}^{2}-|\chi|^{2}}}}\right],
\end{equation}
and
\begin{equation}
\partial_{\theta} t_{i}=-\frac{\nu}{4}\left(\chi_{i} \partial_{\theta} \chi_{0}\right)+\frac{\mu}{2}\left(\partial_{\theta} \chi_{i}\right)+\frac{\gamma}{4}\chi_{i}\sum_{j=1}^{3}\left(\chi_{j} \partial_{\theta} \chi_{j}\right),
\end{equation}
where the quantities $\nu$, $\mu$ and $\gamma$ are given by
\begin{equation}\label{444}
		\nu =\left(\chi_{0}^{2}-|\chi|^{2}\right)^{-1/2} \left(\chi_{0}+\sqrt{\chi_{0}^{2}-|\chi|^{2}}\right)^{-1/2}, \hspace{1cm}
		\mu=\left(\chi_{0}+\sqrt{\chi_{0}^{2}-|\chi|^{2}}\right)^{-1/2},
\end{equation}
and
\begin{equation}
\gamma =\left(\chi_{0}^{2}-|\chi|^{2}\right)^{-1/2} \left(\chi_{0}+\sqrt{\chi_{0}^{2}-|\chi|^{2}}\right)^{-3/2}.
\end{equation}
Consequently, the evaluation of $\partial \tilde{t}_{0}$ and $\partial \tilde{t}_{i}$, as a linear combination of $\partial_{\theta} \tilde{\chi}_{0}$ and  $\partial_{\theta} \tilde{\chi}_{i}$, is required to obtain the skew information associated with the Hermitian matrix $\sqrt{\rho_{\theta}^{(2)}}$. For pure states, the skew information reduces to
\begin{equation}\label{pure I}
		\mathcal{I}\left(\rho_{\theta}^{(1)}\right)=2\left(|\partial_{\theta}\chi_{0}|^{2}+\sum_{i=1}^{3}|\partial_{\theta}\chi_{i}|^{2}\right),\hspace{1cm}{\rm and}\hspace{1cm} \mathcal{I}(\rho_{\theta}^{(2)})=2\left(|\partial_{\theta}\tilde{\chi}_{0}|^{2}+\sum_{i=1}^{3}|\partial_{\theta}\tilde{\chi}_{i}|^{2}\right), 
\end{equation}
in term of $\chi_{0}$, $\chi_{i}$, $\tilde{\chi}_{0}$ and $\tilde{\chi}_{i}$. 
\section{Explicit expressions of quantum Fisher and skew information under decoherence effect}
\subsection{Quantum channels}
Open quantum systems, which are coupled to an external environment, are playing a major role in quantum information theory. In this situation, the efficiency of quantum information processing depends critically on the details of decoherence and noise-induced by the environment \cite{Li2013,Maziero2009}. It is, therefore, necessary to find a concrete mathematical description of the errors that reflects our understanding of the environmental influence \cite{Bourennane2004}. The Markov dynamics of states evolving in noisy environments are modeled by the maps between the spaces of operators known as quantum channels. Indeed, a quantum channel is a map $\Phi$ that acts on the density operators; $\Phi: \rho\longmapsto\Phi\left(\rho\right)$, and incorporates the evolution of a pure quantum state to a mixed quantum state \cite{Schlosshauer2007,Hornberger2009}. Mathematically, the description of the quantum channel can be completely characterized by the Kraus representation as
\begin{equation}
\rho_{\rm out}=\Phi\left(\rho\right)=\sum_{i}\mathcal{K}_{i}\rho\mathcal{K}_{i}^{\dagger},\hspace{1cm}{\rm and}\hspace{1cm}\sum_{i}\mathcal{K}_{i}^{\dagger}\mathcal{K}_{i}=\mathbb{I},
\end{equation}
where $\mathcal{K}_{i}$ are the Kraus operator. For a two-qubit system, described by the density matrix $\rho$, this action can be expressed in the Fano-Bloch representation as
\begin{equation}\label{aa}
	\Phi(\rho)=\sum_{\alpha \beta}(\Phi(\mathcal{T}))_{\alpha \beta} \sigma_{\alpha} \otimes \sigma_{\beta},
\end{equation}
where $\sigma_{\alpha}$ ($\alpha=1,2,3$) are the Pauli matrices and the matrix elements $(\Phi(\mathcal{T}))_{\alpha \beta}$ are given by
\begin{equation}\label{bb}
	\left(\Phi\left(\mathcal{T}\right)\right)_{\alpha \beta}=\operatorname{Tr}\left(\Phi^{\dagger}\left(\sigma_{\alpha}\right) \otimes \Phi^{\dagger}\left(\sigma_{\beta}\right) \rho\right),
\end{equation}
with
\begin{equation}\label{cc}
	\Phi^{\dagger}\left(\sigma_{\alpha}\right)=\sum_{i}\left(\mathcal{K}_{i}\right)^{\dagger} \sigma_{\alpha} \mathcal{K}_{i}.
\end{equation}
 It is clear to check that $\Phi^{\dagger}\left(\sigma_{\alpha}\right)$ reads as
\begin{equation}\label{dd}
	\Phi^{\dagger}\left(\sigma_{\alpha}\right)=\sum_{\alpha^{\prime}} W_{\alpha \alpha^{\prime}} \sigma_{\alpha^{\prime}},
\end{equation}
where the matrix elements of the transformation $W$ are given by
\begin{equation}\label{ee}
	W_{\alpha \alpha^{\prime}}=\frac{1}{2} \operatorname{Tr}\left(\Phi^{\dagger}\left(\sigma_{\alpha}\right) \sigma_{\alpha^{\prime}}\right).
\end{equation}
Using the relations (\ref{bb}) and (\ref{dd}), one obtains
\begin{equation} \label{ff}
	\Phi(\mathcal{T})=W \mathcal{T} W^{T}.
\end{equation}
where we denote the transpose of matrix $W$ by $W^{T}$. With the above description, we will focus on the dynamics of the Fisher and skew information for two-qubit system under noisy channels (i.e., phase-damping, depolarizing and amplitude-damping channels), which can be modeled in terms of Kraus operators as follows:\\
{\bf ($i$) Phase-damping channel:} The phase-damping channel describes the loss of quantum information without loss of energy. The corresponding Kraus operators are given by
\begin{equation}\label{PDC}
	\mathcal{K}_{0}^{PDC}=\sqrt{1-p} \mathbb{I}, \hspace{1cm} \mathcal{K}_{1}^{PDC}=\sqrt{\frac{p}{4}}(\mathbb{I}+\sigma_{z}), \hspace{1cm} {\rm and}\hspace{1cm} \mathcal{K}_{2}^{PDC}=\sqrt{\frac{p}{4}}(\mathbb{I}-\sigma_{z}).
\end{equation}
where the decoherence probability $p=1-s=1-e^{-\gamma t}$ with $\gamma$ is the decay rate.\\
{\bf ($ii$) Depolarizing channel:} The depolarizing channel is a model for decoherence where with a certain probability of $3p/4$ the qubit becomes totally mixed. The three Kraus operators are characterized by the set
\begin{equation}\label{DPC}
	\mathcal{K}_{0}^{DPC}=\sqrt{1-\frac{3p}{4}}\mathbb{I}, \hspace{1cm} \mathcal{K}_{1}^{DPC}=\sqrt{\frac{p}{4}}\sigma_{x} , \hspace{1cm} \mathcal{K}_{2}^{DPC}=\sqrt{\frac{p}{4}}\sigma_{y}, \hspace{1cm} {\rm and}\hspace{1cm} \mathcal{K}_{3}^{DPC}=\sqrt{\frac{p}{4}}\sigma_{z}.
\end{equation}
{\bf ($iii$) Amplitude-damping channel:} The amplitude damping channel can be used to describe, for example, the disintegration of a two-level excited state of an atom due to the spontaneous emission of a photon. Obviously, the Kraus operators are given by
\begin{equation}\label{ADC}
	\mathcal{K}_{1}^{ADC}=\sqrt{s}|0\rangle\langle 0|+| 1\rangle\left\langle 1\left|, \hspace{1cm} {\rm and} \hspace{1cm} \mathcal{K}_{2}^{ADC}=\sqrt{p}\right| 1\right\rangle\langle 0|.
\end{equation}
These quantum channels are the prototype models of dissipation relevant in various experimental systems. Especially, they provide a revealing caricature of decoherence in realistic physical situations, with all inessential mathematical details stripped away. Some of the examples include multiphoton systems, ion traps, atomic ensembles, or solid-state spin systems such as quantum dots.
\subsection{Fisher and skew information for evolved two-qubit $X$-states}
In this section, we shall discuss the effects of these three different noisy channels on both the quantum Fisher information and the skew information for two-qubit $X$-states. Clearly, under decoherence effect, the evolved density matrix $\Phi(\varrho)$ remains a two-qubit $X$-state. Using the linearity property of quantum channels, i.e., $\Phi\left(\alpha\rho_{1}+\beta\rho_{2}\right)=\alpha\Phi\left(\rho_{1} \right)+\beta\Phi\left(\rho_{2}\right)$, and the decomposition of equation (\ref{D}), we obtain
\begin{equation}\label{ii}
	\Phi(\rho_{AB})=\Phi(\rho_{AB}^{(1)})+\Phi(\rho_{AB}^{(2)}),
\end{equation}
where $\Phi(\rho_{AB}^{(1)})$ and $\Phi(\rho_{AB}^{(2)})$ given by
\begin{equation}\label{jj}
	\Phi(\rho_{AB}^{(1)})=\frac{1}{2} \sum_{\alpha=0}^{3} \Lambda_{\alpha} \vartheta_{\alpha},
	\hspace{1cm}{\rm and}\hspace{1cm} \Phi(\rho_{AB}^{(2)})=\frac{1}{2} \sum_{\alpha=0}^{3} \tilde{\Lambda}_{\alpha} \tilde{\vartheta}_{\alpha}.
\end{equation}
Therefore, the quantum Fisher information for the state (\ref{ii}) can be written as
\begin{equation}\label{50}
	\mathcal{F}\left(\Phi(\rho_{AB}) \right)=\mathcal{F}\left(\Phi(\rho_{AB}^{(1)})\right) +\mathcal{F}\left(\Phi(\rho_{AB}^{(2)}) \right),
\end{equation}
where
\begin{equation}\label{51}
\mathcal{F}\left(\Phi(\rho_{AB}^{(1)})\right)=\frac{\left(\partial_{\theta} \Lambda_{0}\right)^{2}}{\Lambda_{0}}+\frac{1}{\Lambda_{0}}\left[\frac{\left(g_{\alpha \beta} \Lambda^{\alpha} \partial_{\theta} \Lambda^{\beta}\right)^{2}}{g_{\alpha \beta} \Lambda^{\alpha} \Lambda^{\beta}}-g_{\alpha \beta}\left(\partial_{\theta} \Lambda^{\alpha}\right)\left(\partial_{\theta} \Lambda^{\beta}\right)\right],
\end{equation}
and
\begin{equation}
\mathcal{F}\left(\Phi(\rho_{AB}^{(2)})\right)=\frac{\left(\partial_{\theta}\tilde{\Lambda}_{0}^{2}\right)}{\tilde{\Lambda}_{0}}+\frac{1}{\tilde{\Lambda}_{0}}\left[\frac{\left(g_{\alpha \beta} \tilde{\Lambda}^{\alpha} \partial_{\theta} \tilde{\Lambda}^{\beta}\right)^{2}}{g_{\alpha \beta} \tilde{\Lambda}^{\alpha} \tilde{\Lambda}^{\beta}}-g_{\alpha \beta}\left(\partial_{\theta} \tilde{\Lambda}^{\alpha}\right)\left(\partial_{\theta} \tilde{\Lambda}^{\beta}\right)\right].
\end{equation}
For pure states, the quantum Fisher information reduces to the following explicit form
\begin{equation}\label{pure D1}
	\mathcal{F}\left(\Phi(\rho_{AB}^{(1)}) \right)=(\Lambda_{0})^{2}+\sum_{i=1}^{3}(\Lambda_{i})^{2},\hspace{1cm}{\rm}\hspace{1cm}{\rm and}\hspace{1cm}\mathcal{F}\left(\Phi(\rho_{AB}^{(2)}) \right)=(\tilde{\Lambda_{0}})^{2}+\sum_{i=1}^{3}(\tilde{\Lambda_{i}})^{2}.
\end{equation}
As we saw earlier, for the noisy effects under consideration, any two-qubit $X$-state remains $X$-shaped. Further, the states $\sqrt{\Phi(\rho_{AB}^{(j)})}$ (with $j=1,2$) can be written as a linear combination of the generators $\vartheta_{\alpha}$ and $\tilde{\vartheta}_{\alpha}$ as
\begin{equation}
\sqrt{\Phi(\rho_{AB}^{(1)})}=\frac{1}{2}\sum_{\alpha=0}^{3}X_{\alpha}\vartheta_{\alpha},
\hspace{1cm}{\rm and}\hspace{1cm}\sqrt{\Phi(\rho_{AB}^{(2)})}=\frac{1}{2}\sum_{\alpha=0}^{3}\tilde{X}_{\alpha}\tilde{\vartheta}_{\alpha},
\end{equation}
where $X_{\alpha}$ (resp.$\tilde{X}_{\alpha}$) are the components of the Bloch vector of the matrix $\sqrt{\Phi(\rho_{AB}^{(1)})}$ (resp.$\sqrt{\Phi(\rho_{AB}^{(2)})}$). Consequently, it is easy to see that skew information is written as
\begin{equation}\label{52}
\mathcal{I}\left(\Phi(\rho_{AB}^{(1)}) \right)=8\left[\left(\partial_{\theta} X_{0}\right)^{2}+\sum_{i=1}^{3}\left(\partial_{\theta} X_{i}\right)^{2}\right], \hspace{1cm}{\rm and}\hspace{1cm}\mathcal{I}\left(\Phi(\rho_{AB}^{(2)}) \right)=8\left[\left(\partial \tilde{X}_{0}\right)^{2}+\sum_{i=1}^{3}\left(\partial \tilde{X}_{i}\right)^{2}\right].
\end{equation}
For pure states, the skew information is expressed as below:
\begin{equation}\label{pure D2}
\mathcal{I}\left(\Phi(\rho_{AB}^{(1)}) \right)=2\left(|\partial_{\theta}\Lambda_{0}|^{2}+\sum_{i=1}^{3}|\partial_{\theta}\Lambda_{i}|^{2}\right), \hspace{1cm} {\rm and}\hspace{1cm}\mathcal{I}\left(\Phi(\rho_{AB}^{(2)}) \right)=2\left(|\partial_{\theta}\tilde{\Lambda}_{0}|^{2}+\sum_{i=1}^{3}|\partial_{\theta}\tilde{\Lambda}_{i}|^{2}\right).
\end{equation}
Thus, the analytical expression of quantum Fisher and skew information are obtained for any two-qubit system by deriving the correlation matrix $\Phi(\mathcal{T})$. For phase-damping channel, which is defined by the set of Kraus operators (\ref{PDC}), the action of the application $\Phi$ on the correlation matrix $\mathcal{T}$ is
\begin{equation}\label{phiT1}
	\Phi(\mathcal{T})=\left(\begin{array}{cccc}
		\mathcal{T}_{00} & 0 & 0 & \mathcal{T}_{03} \\
		0 & s^{2} \mathcal{T}_{11} & s^{2} \mathcal{T}_{12} & 0 \\
		0 & s^{2} \mathcal{T}_{21} & s^{2} \mathcal{T}_{22} & 0 \\
		\mathcal{T}_{30} & 0 & 0 & \mathcal{T}_{33}
	\end{array}\right).
\end{equation}
and the quantities $\Lambda_{i}$ and $\tilde{\Lambda}_{i}$ are, respectively, given by
\begin{equation}\label{47}
		\Lambda_{0}=\chi_{0}, \hspace{1cm} \Lambda_{1}=s^2\chi_{1}, \hspace{1cm} \Lambda_{2}=s^2\chi_{2}, \hspace{1cm} \Lambda_{3}=\chi_{3},
\end{equation}
and
\begin{equation}
\tilde{\Lambda}_{0}=\tilde{\chi}_{0}, \hspace{1cm} \tilde{\Lambda}_{1}=s^2\tilde{\chi}_{1}, \hspace{1cm} \tilde{\Lambda}_{2}=s^2\tilde{\chi}_{2}, \hspace{1cm} \tilde{\Lambda}_{3}=\tilde{\chi}_{3}.
\end{equation}
Likewise, it is simple to verify that the matrix (\ref{ff}) for the depolarizing channel is expressed as
\begin{equation}\label{phiT2}
	\Phi(\mathcal{T})=\left(\begin{array}{cccc}
		\mathcal{T}_{00} & 0 & 0 & s\mathcal{T}_{03} \\
		0 & s^{2} \mathcal{T}_{11} & s^{2} \mathcal{T}_{12} & 0 \\
		0 & s^{2} \mathcal{T}_{21} & s^{2} \mathcal{T}_{22} & 0 \\
		s\mathcal{T}_{30} & 0 & 0 & s^{2}\mathcal{T}_{33}
	\end{array}\right),
\end{equation}
and in such a case, we obtain
\begin{equation}
\Lambda_{0}=\frac{1}{2}((\chi_{0}+\tilde{\chi}_{0})+s^{2}(\chi_{0}-\tilde{\chi}_{0})), \hspace{1cm} \Lambda_{1}=s^2\chi_{1}, \hspace{1cm} \Lambda_{2}=s^2\chi_{2}, \hspace{1cm} \Lambda_{3}=s^2\chi_{3},
\end{equation}
and
\begin{equation}
\tilde{\Lambda}_{0}=\frac{1}{2}((\chi_{0}+\tilde{\chi}_{0})-s^{2}(\chi_{0}-\tilde{\chi}_{0})), \hspace{1cm} \tilde{\Lambda}_{1}=s^2\chi_{1},\hspace{1cm} \tilde{\Lambda}_{2}=s^2\chi_{2},\hspace{1cm} \tilde{\Lambda}_{3}=s^2\chi_{3}.
\end{equation}
Further, through the Kraus operator (\ref{ADC}) that characterizes the amplitude damping channel, we prove that
\begin{equation}\label{phiT3}
	\Phi(\mathcal{T})=\left(\begin{array}{cccc}
		\mathcal{T}_{00} & 0 & 0 & s\mathcal{T}_{03}-p\mathcal{T}_{00} \\
		0 & s \mathcal{T}_{11} & s \mathcal{T}_{12} & 0 \\
		0 & s \mathcal{T}_{21} & s \mathcal{T}_{22} & 0 \\
		s\mathcal{T}_{30}-p\mathcal{T}_{00} & 0 & 0 & p^{2}\mathcal{T}_{00}-sp(\mathcal{T}_{30}+\mathcal{T}_{03})+s^{2}\mathcal{T}_{33}
	\end{array}\right),
\end{equation}
where
\begin{equation}
\Lambda_{0}=\frac{1}{2}\left[(1+p^{2}+s^{2})\chi_{0}-2sp\chi_{3}+(1+p^{2}-s^{2})\tilde{\chi}_{0}\right], \quad \Lambda_{1}=s\chi_{1}, \hspace{1cm} \Lambda_{2}=s\chi_{2}, \quad \Lambda_{3}=s\chi_{3}-p(\chi_{0}+\tilde{\chi}_{0}),
\end{equation}
and
\begin{equation}
\tilde{\Lambda}_{0}=\frac{1}{2}\left[(1-p^{2}-s^{2})\chi_{0}+2sp\chi_{3}+(1-p^{2}+s^{2})\tilde{\chi}_{0}\right], \quad \tilde{\Lambda}_{1}=s\tilde{\chi}_{1}, \hspace{1cm} \tilde{\Lambda}_{2}=s\tilde{\chi}_{2}, \hspace{1cm} \tilde{\Lambda}_{3}=s\tilde{\chi}_{3},
\end{equation}
Clearly, the correlation matrix $\Phi(\mathcal{T})$ given by (\ref{phiT1}), (\ref{phiT2}) and (\ref{phiT3}) corresponds to an $X$-shaped density matrix. Therefore, the three noisy channels preserves the $X$-form of the initial $X$-state. In this respect, the results obtained in the previous section can be adopted to investigate the quantum Fisher and skew information in three different noisy channels for the system under consideration.

\section{Application to quasi-Werner states}
Coherent states \cite{Monroe1996,Zhang1990}, which were discussed by Glauber \cite{Glauber1963} to explain the coherence property of laser light, are considered the closest quantum counterpart to classical radiation field states. They have an important role in quantum information processing to storage, process or teleport an unknown quantum state. Actually, the coherent states are eigenstates of the annihilation operator $a$, i.e., $a\left|\alpha \right\rangle=\alpha\left|\alpha \right\rangle$ with the Fock state representation of a single mode coherent state is
\begin{equation}
\left|\alpha \right\rangle=D\left(\alpha\right)\left|0\right\rangle =\exp\left[-\frac{1}{2}\mid\alpha\mid^{2}\right]\sum_{n}\frac{\alpha^{n}}{\sqrt{n!}}\left|n\right\rangle,\label{CS}
\end{equation}
which can be viewed as the unitary displacement action $D\left(\alpha\right)$ on the ground state vector $\left|0\right\rangle$, with
\begin{equation}
D\left(\alpha\right)=\exp\left[-\mid\alpha\mid^{2}/2\right]\exp\left[\alpha a^{\dagger}\right]\exp\left[-\alpha^{*} a\right]. 
\end{equation}
Besides, the coherent state (\ref{CS}) is a typical example of the non-orthogonal state, so it's easy to see that
\begin{equation}
\left\langle\alpha_{1} \mid\alpha_{2}\right\rangle=\exp\left[-\frac{1}{2}\mid\alpha_{1}\mid^{2}-\frac{1}{2}\mid\alpha_{2}\mid^{2}+\alpha_{1}^{*}\alpha_{2} \right]. 
\end{equation}
Moreover, the coherent state family solves unity and allows the over completeness in the space such that
\begin{equation}
\frac{1}{\pi}\int\left|\alpha\right\rangle \left\langle\alpha\right| d^{2}\alpha=\openone, \hspace{1cm}{\rm and}\hspace{1cm}d^{2}\alpha=d\mathcal{R}e\left(\alpha\right)+d\mathcal{I}m\left(\alpha\right).
\end{equation} 
If we consider the two coherent states $\left|\alpha \right\rangle$ and $\left|\beta\right\rangle$ with corresponding coherent amplitudes $\alpha$ and $\beta$, we can write the two bipartite superposed coherent states as
\begin{equation}\label{1}
	\left|\psi^{\pm}\right\rangle=n_{\pm}\left[\left|\alpha, \beta\right\rangle \pm\left|-\alpha,-\beta\right\rangle\right],
\end{equation}
where the states $\left|-\alpha \right\rangle$ and $\left|-\beta\right\rangle$ are $\pi$-radian out of phase with the corresponding coherent states and $n_{\pm}$ is the normalization factor reads
\begin{equation}\label{2}
	n_{\pm}=\left[2\left(1 \pm \chi_{\alpha}^{2} \chi_{\beta}^{2}\right)\right]^{-\frac{1}{2}},
\end{equation}
with $\chi_{\alpha}=\exp\left[-|\alpha|^{2}\right]$ and $\chi_{\beta}=\exp\left[-|\beta|^{2}\right]$. As mentioned earlier, the coherent states of the form (\ref{CS}) are not orthogonal to each other, however, an orthonormal basis can be achieved by considering the even and odd coherent states $\mid\pm_{\alpha}\rangle$ and $\mid\pm_{\beta}\rangle$ as
\begin{equation}\label{3}
		\left|\pm_{\alpha}\right\rangle =N_{\pm}^{\alpha}[|\alpha\rangle \pm|-\alpha\rangle], \hspace{1cm}{\rm and}\hspace{1cm}
		\left|\pm_{\beta}\right\rangle =N_{\pm}^{\beta}[|\beta\rangle \pm|-\beta\rangle],
\end{equation}
where the normalization constants are given by
\begin{equation}\label{4}
		N_{\pm}^{\alpha}=\left[2\left(1 \pm \chi_{\alpha}^{2}\right)\right]^{-\frac{1}{2}},\hspace{1cm}{\rm and}\hspace{1cm} N_{\pm}^{\beta}=\left[2\left(1 \pm \chi_{\beta}^{2}\right)\right]^{-\frac{1}{2}}.
\end{equation}
Reporting the expressions (\ref{3}) and (\ref{4}) in equation (\ref{1}), we rewrite the states $\left|\psi^{\pm}\right\rangle$ using the orthonormal basis states as follows
\begin{equation}\label{5}
	\left|\psi^{+}\right\rangle=\frac{n^{+}}{2}\left[\frac{\left|+_{\alpha},+_{\beta}\right\rangle}{N_{+}^{\alpha} N_{+}^{\beta}}+\frac{|-_{\alpha},-_{\beta}\rangle}{N_{-}^{\alpha} N_{-}^{\beta}}\right], \hspace{1cm}{\rm and}\hspace{1cm}
		\left|\psi^{-}\right\rangle=\frac{n^{-}}{2}\left[\frac{\left|+_{\alpha},-_{\beta}\right\rangle}{N_{+}^{\alpha} N_{-}^{\beta}}+\frac{|-_{\alpha},+_{\beta}\rangle}{N_{-}^{\alpha} N_{+}^{\beta}}\right].
\end{equation}
Respectively, two quasi-Werner states \cite{Werner1989} based on the bipartite superposed coherent states are written as
\begin{equation}\label{6}
	\rho(\psi^{+},q)=(1-q)\frac{\openone}{4}+q|\psi^{+}\rangle\langle\psi^{+}|, \hspace{1cm}{\rm and}\hspace{1cm} \rho(\psi^{-},q)=(1-q)\frac{\openone}{4}+q|\psi^{-}\rangle\langle\psi^{-}|,
\end{equation}
where $q$ being the mixing parameter ranging from $0$ (for maximal mixed states) to $1$ (for pure states). The density matrices $\rho\left(\psi^{\pm}, q\right)$, associated to the quasi-Werner states $\left|\psi^{\pm}\right\rangle$, take the following forms
\begin{equation}\label{+}
		\rho\left(\psi^{+}, p\right)=\frac{1}{4}\left(\begin{array}{cccc}
			1+q\left(\frac{n_{+}^{2}}{\left(N_{+}^{\alpha}\right)^{2}\left(N_{+}^{\beta}\right)^{2}}-1\right) & 0 & 0 & \frac{q n_{+}^{2}}{N_{+}^{\alpha} N_{+}^{\beta} N_{-}^{\alpha} N_{-}^{\beta}} \\
			0 & 1-q & 0 & 0 \\
			0 & 0 & 1-q & 0 \\
			\frac{q n_{+}^{2}}{ N_{+}^{\alpha} N_{+}^{\beta} N_{-}^{\alpha} N_{-}^{\beta}} & 0 & 0 & 1+q\left(\frac{n_{+}^{2}}{\left(N_{-}^{\alpha}\right)^{2}\left(N_{-}^{\beta}\right)^{2}}-1\right)
		\end{array}\right),
\end{equation}
and
\begin{equation}
\rho\left(\psi^{-},q\right)=\frac{1}{4}\left(\begin{array}{cccc}
	1-q & 0 & 0 & 0 \\
	0 & 1+q\left(\frac{n_{-}^{2}}{\left(N_{+}^{\alpha}\right)^{2}\left(N_{-}^{\beta}\right)^{2}}-1\right) & \frac{qn_{-}^{2}}{ N_{+}^{\alpha} N_{+}^{\beta} N_{-}^{\alpha} N_{-}^{\beta}} & 0 \\
	0 & \frac{qn_{-}^{2}}{ N_{+}^{\alpha} N_{+}^{\beta} N_{-}^{\alpha} N_{-}^{\beta}} & 1+q\left(\frac{n_{-}^{2}}{\left(N_{-}^{\alpha}\right)^{2}\left(N_{+}^{\beta}\right)^{2}}-1\right) & 0 \\
	0 & 0 & 0 & 1-q
\end{array}\right),\label{-}
\end{equation}
in the computational basis
$\{|+_\alpha,+_\beta\rangle, |+_\alpha,-_\beta\rangle,
|-_\alpha,+_\beta\rangle,|-_\alpha,-_\beta\rangle\}$. The non-vanishing matrix elements $\mathcal{T}_{\alpha\beta}$ (with $\alpha,\beta=0,1,2,3$) are given by
\begin{equation}
	\begin{aligned}
		&\mathcal{T}_{00}^{+}=1, \hspace{1cm} \mathcal{T}_{11}^{+}=\frac{-qn_{+}^{2}}{2N_{+}^{\alpha}N_{+}^{\beta}N_{-}^{\alpha}N_{-}^{\beta}}, \hspace{1cm} \mathcal{T}_{30}^{+}=\frac{qn_{+}^{2}}{4}\left(\frac{1}{(N_{\alpha}^{+}N_{\beta}^{+})^{2}}-\frac{1}{(N_{\alpha}^{-}N_{\beta}^{-})^{2}}\right), \\
		&\quad \mathcal{T}_{33}^{+}=q , \hspace{1cm} \mathcal{T}_{22}^{+}=\frac{qn_{+}^{2}}{2N_{+}^{\alpha}N_{+}^{\beta}N_{-}^{\alpha}N_{-}^{\beta}} , \hspace{1cm} \mathcal{T}_{03}^{+}=\frac{qn_{+}^{2}}{4}\left(\frac{1}{(N_{\alpha}^{+}N_{\beta}^{+})^{2}}-\frac{1}{(N_{\alpha}^{-}N_{\beta}^{-})^{2}}\right),
	\end{aligned}
\end{equation}
and
\begin{equation}
	\begin{aligned}
		&\mathcal{T}_{00}^{-}=1, \hspace{1cm} \mathcal{T}_{11}^{-}=\frac{qn_{-}^{2}}{2N_{+}^{\alpha}N_{+}^{\beta}N_{-}^{\alpha}N_{-}^{\beta}}, \hspace{1cm} \mathcal{T}_{30}^{-}=\frac{qn_{-}^{2}}{4}\left(\frac{1}{(N_{\alpha}^{+}N_{\beta}^{-})^{2}}-\frac{1}{(N_{\alpha}^{-}N_{\beta}^{+})^{2}}\right), \\
		&\quad \mathcal{T}_{33}^{-}=-q ,\hspace{1cm} \mathcal{T}_{22}^{-}=\frac{qn_{-}^{2}}{2N_{+}^{\alpha}N_{+}^{\beta}N_{-}^{\alpha}N_{-}^{\beta}} , \hspace{1cm} \mathcal{T}_{03}^{-}=\frac{-qn_{-}^{2}}{4}\left(\frac{1}{(N_{\alpha}^{+}N_{\beta}^{-})^{2}}-\frac{1}{(N_{\alpha}^{-}N_{\beta}^{+})^{2}}\right),
	\end{aligned}
\end{equation}
for the states $\rho\left(\psi^{+}, q\right)$ and $\rho\left(\psi^{-}, q\right)$, respectively. Then we decompose these density matrices as
\begin{equation}
\rho\left(\psi^{\pm},q\right)\equiv\rho_{\pm}=\varrho_{1}^{\pm}+\varrho_{2}^{\pm},
\end{equation}
where
\begin{equation}
	\begin{aligned}
		&\varrho_{1}^{\pm}=\frac{1}{2}(\chi_{0}^{\pm}\vartheta_{0}+\chi^{\pm}\vartheta_{1}),
		\hspace{1cm}{\rm and}\hspace{1cm} \varrho_{2}^{\pm}=\frac{1}{2}(\tilde{\chi}_{0}^{\pm}\tilde{\vartheta}_{0}+\tilde{\chi}_{1}^{\pm}\tilde{\vartheta}_{1}).
	\end{aligned}
\end{equation}
The non-vanishing quantities $\chi_{i}^{\pm}$ and $\tilde{\chi}_{i}^{\pm}$ are given by
\begin{equation}
	\begin{aligned}
		&\chi_{0}^{\pm}=\frac{1\pm q}{2}, \hspace{1cm} \chi_{1}^{+}=\frac{qn_{+}^{2}}{2N_{+}^{\alpha}N_{+}^{\beta}N_{-}^{\alpha}N_{-}^{\beta}}, \hspace{1cm} \chi_{3}^{+}=\frac{qn_{+}^{2}}{4}\left(\frac{1}{(N_{\alpha}^{+}N_{\beta}^{+})^{2}}-\frac{1}{(N_{\alpha}^{-}N_{\beta}^{-})^{2}}\right),  \\
		&\tilde{\chi}_{0}^{\pm}=\frac{1\mp q}{2}, \hspace{1cm} \tilde{\chi}_{1}^{-}=\frac{qn_{-}^{2}}{2N_{+}^{\alpha}N_{+}^{\beta}N_{-}^{\alpha}N_{-}^{\beta}},\hspace{1cm} \tilde{\chi}_{3}^{-}=\frac{qn_{-}^{2}}{4}\left(\frac{1}{(N_{\alpha}^{+}N_{\beta}^{-})^{2}}-\frac{1}{(N_{\alpha}^{-}N_{\beta}^{+})^{2}}\right).
	\end{aligned}
\end{equation}
In order to investigate the role of quantum entanglement in quantum metrology for the bipartite quasi-Werner states, we employ the most accepted bona fide entanglement measure called concurrence ${\cal C}\left(\rho\right)$, which was introduced by Wootters \cite{Wootters1998}. For our states $\rho\left(\psi^{\pm},q\right)$, the analytical expression of concurrence takes the form 
\begin{equation}\label{Con}
	{\cal C}\left(\rho_{\pm}\right)=\max\left\lbrace0, \frac{qn_{\pm}^{2}-\left(1-q\right)N_{+}^{\alpha} N_{-}^{\alpha} N_{+}^{\beta} N_{-}^{\beta}}{2N_{+}^{\alpha} N_{-}^{\alpha} N_{+}^{\beta} N_{-}^{\beta}}\right\rbrace.
\end{equation}
\begin{figure}[h!]
	{{\begin{minipage}[b]{.5\linewidth}
				\centering
				\includegraphics[scale=0.45]{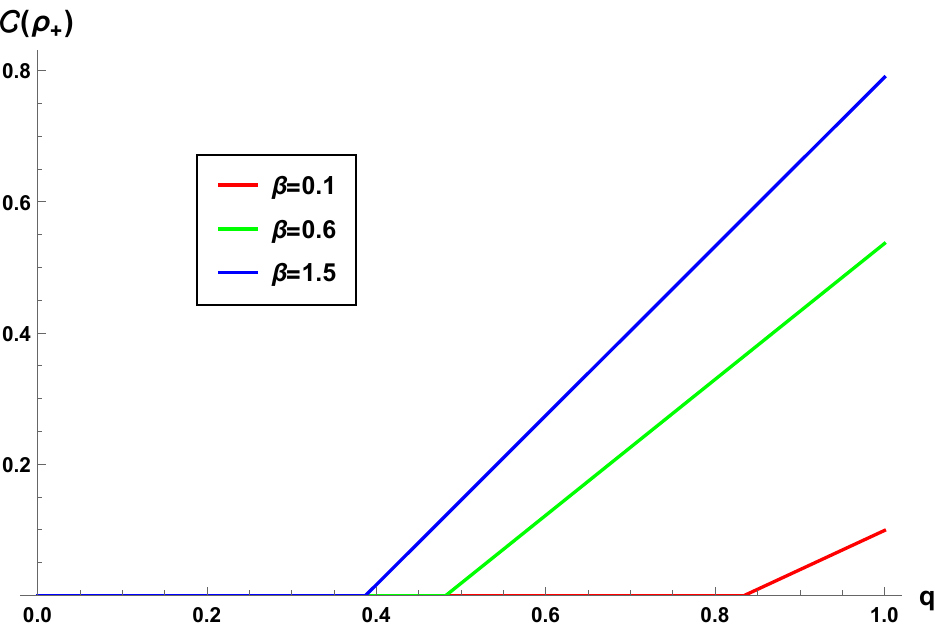}
			\end{minipage}\hfill
			\begin{minipage}[b]{.5\linewidth}
				\centering
				\includegraphics[scale=0.45]{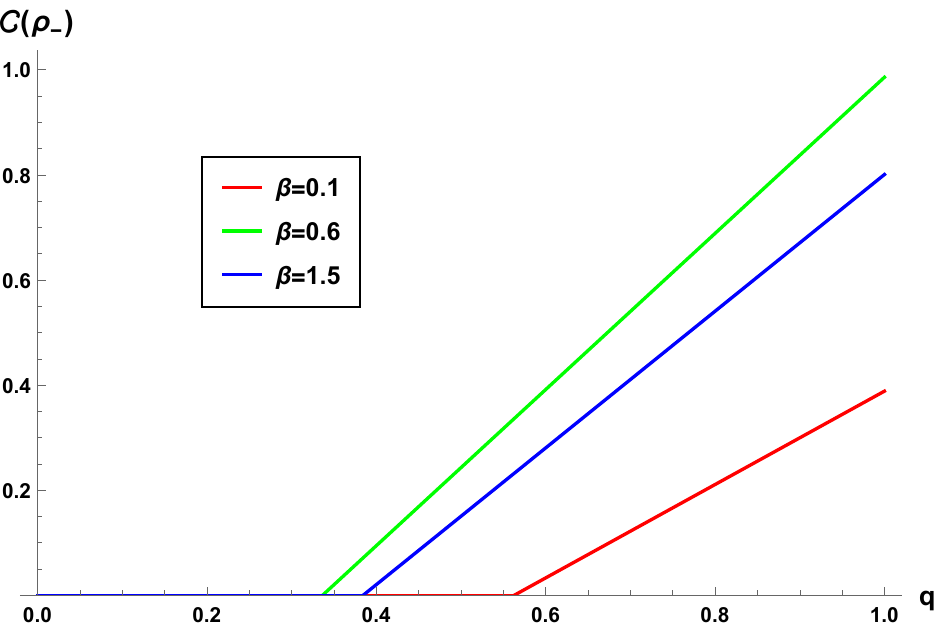}
	\end{minipage}}}
	\caption{Variation of the concurrence ${\cal C}\left(\rho_{\pm}\right)$ versus the mixing parameter $q$ for the different values of the coherent amplitude $\beta$ with $\alpha=0.5$.}\label{Fig1}
\end{figure}
To get an insight into the influence of the mixedness on the entanglement dynamics of two quasi-Werner states, we display the concurrence (\ref{Con}) as a function of the mixing paramete $q$ for several values of the coherent amplitudes $\beta$ in Fig.\ref{Fig1} when $\alpha=0.5$. Based on these results, we observe that the concurrence vanishes for the maximal mixed quasi-Werner states and then increases to achieve its maximum for the pure quasi-Werner states. Moreover, the quantum entanglement increase with increasing values of the coherent amplitudes $\alpha$ for fixed values of mixing parameter $q$. Also, for smaller values of the mixing parameter, the quasi-Werner states $\rho\left(\psi^{\pm},q\right)$ are more separable. This confirms that the mixedness strongly influences the quantum entanglement existing in the quantum system, thus the pure states having always the maximum quantity of quantum correlations.
\subsection{Dynamics of quantum criteria under phase damping channel}
Under the phase damping effect, the evolved matrix density denoted by $\Phi\left(\rho_{\pm}\right)=\rho_{\pm}^{\rm PDC}$ remains $X$-type. Thus, quantum criteria such as quantum Fisher information, concurrence entanglement and skew information can be evaluated using the results reported in the previous section. Using the Kraus operators given by (\ref{PDC}) and after some algebra, it is straightforward to verify that the analytical expression for the quantum Fisher information of the two quasi-Werner states under noisy phase damping reads as follows
\begin{equation}
\mathcal{F}\left(\rho_{\pm}^{\rm PDC} \right)=\frac{3+\left(2q-1\right)\gamma^{\pm}}{2\left(1-q\right)\left(1+2q+q^{2}\gamma^{\pm}\right)},
\end{equation}
where
\begin{equation}
	\gamma^{\pm}=1-\frac{s^{4}n_{\pm}^{4}}{\left(N_{\alpha}^{+}N_{\beta}^{+}N_{\alpha}^{-}
		N_{\beta}^{-}\right)^{2}}-4\kappa_{\pm}^{2},\hspace{1cm}{\rm and}\hspace{1cm}\kappa_{\pm}=\frac{n_{\pm}^{2}\left(N_{\alpha}^{-2}N_{\beta}^{\mp2}-N_{\alpha}^{+2}N_{\beta}^{\pm2} \right)}{4N_{\alpha}^{-2}N_{\beta}^{-2}N_{\alpha}^{+2}N_{\beta}^{+2}}.
\end{equation}
Subsequently, the Wootters concurrence of the two quasi-Werner states under phase damping effect is obtained as
\begin{equation}
	{\cal C}\left(\rho_{\pm}^{\rm PDC} \right)=\max \left\lbrace 0,\frac{s^{2}qn_{\pm}^{2}}{2N_{+}^{\alpha} N_{-}^{\alpha} N_{+}^{\beta} N_{-}^{\beta}}-\frac{1-q}{2}\right\rbrace.
\end{equation}
Based on the above formalism, the analytical expression for the skew information associated with the density matrix $\rho_{\pm}^{\rm PDC}$ is written as
\begin{align}\label{sp1}
	\mathcal{I}\left(\rho_{\pm}^{\rm PDC} \right)&=\frac{1}{2\left(1-q\right)}+\frac{A_{\pm}^{2}}{8}\left(B_{\pm}^{-2}-\frac{qs^{4}n_{\pm}^{4}}{2N_{\alpha}^{+2}N_{\beta}^{+2}N_{\alpha}^{-2}N_{\beta}^{-2}}+2q\kappa_{\pm}^{2}\right)^{2}\notag\\&+\left(2\kappa_{\pm}^{2}+\frac{s^{4}n_{\pm}^{4}}{2N_{\alpha}^{+2}N_{\beta}^{+2}N_{\alpha}^{-2}N_{\beta}^{-2}} \right) \left(B_{\pm}-\frac{qA_{\pm}}{4}+\frac{s^{4}q^{2}n_{\pm}^{4}C_{\pm}}{8N_{\alpha}^{+2}N_{\beta}^{+2}N_{\alpha}^{-2}N_{\beta}^{-2}}+\frac{q^{2}C_{\pm}\kappa_{\pm}^{2}}{2}\right)^{2},
\end{align}
with the entries $A_{\pm}$, $B_{\pm}$ and $C_{\pm}$ in the above equation are given by
\begin{equation}
B_{\pm}=A_{\pm}\omega_{\pm}=\sqrt{2}\left(1+q+\omega_{\pm}\right)^{-1/2},
\end{equation}
and
\begin{equation}
C_{\pm}=B_{\pm}^{3}\omega_{\pm}^{-1},\quad \omega_{\pm}=\sqrt{1+2q+q^{2}\gamma^{\pm}}.
\end{equation}

\begin{figure}[h!]
	{{\begin{minipage}[b]{.33\linewidth}
				\centering
				\includegraphics[scale=0.36]{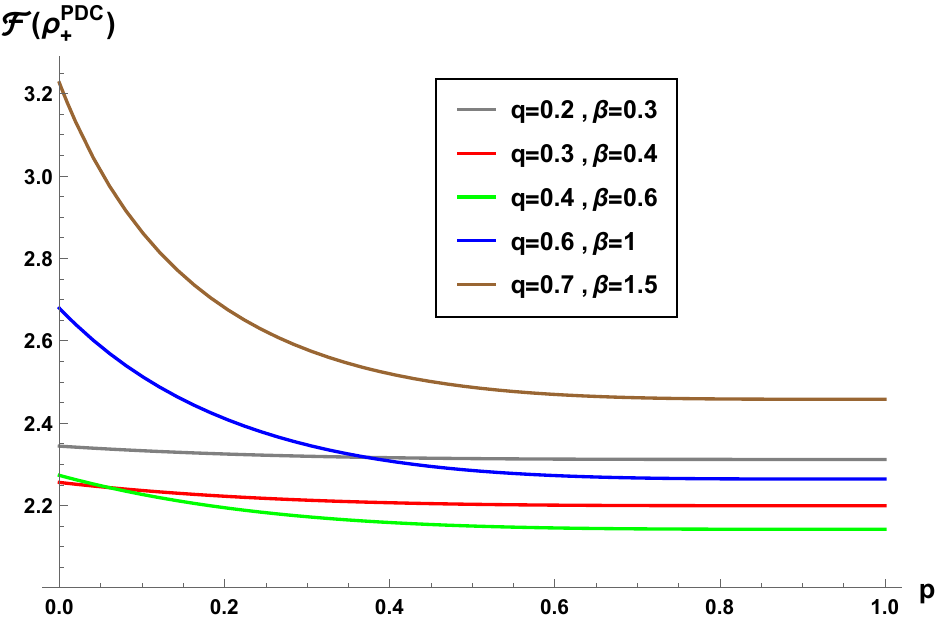}\vfill
				$(a)$
			\end{minipage}\hfill
			\begin{minipage}[b]{.33\linewidth}
				\centering
				\includegraphics[scale=0.36]{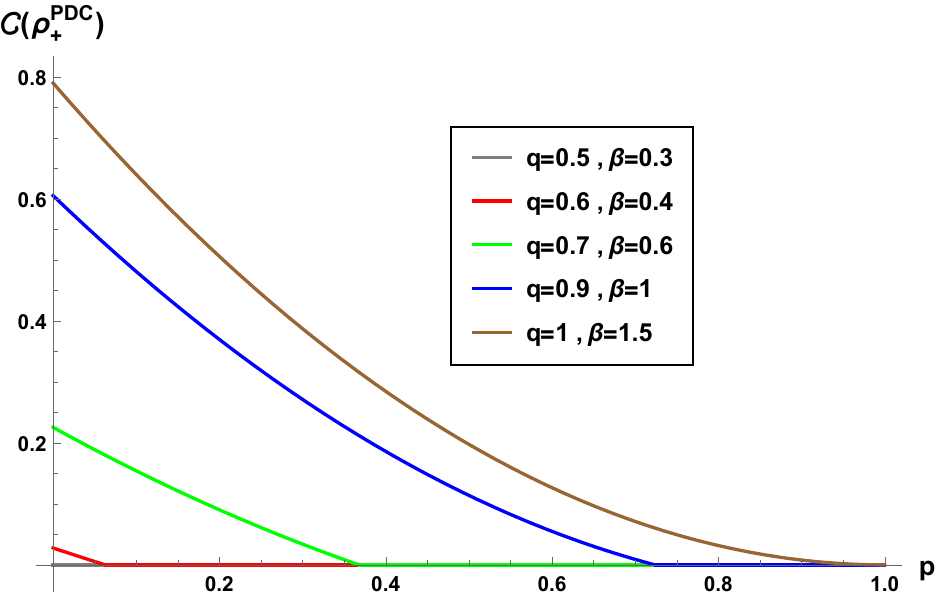}\vfill
				$(b)$
			\end{minipage}\hfill
			\begin{minipage}[b]{.33\linewidth}
				\centering
				\includegraphics[scale=0.36]{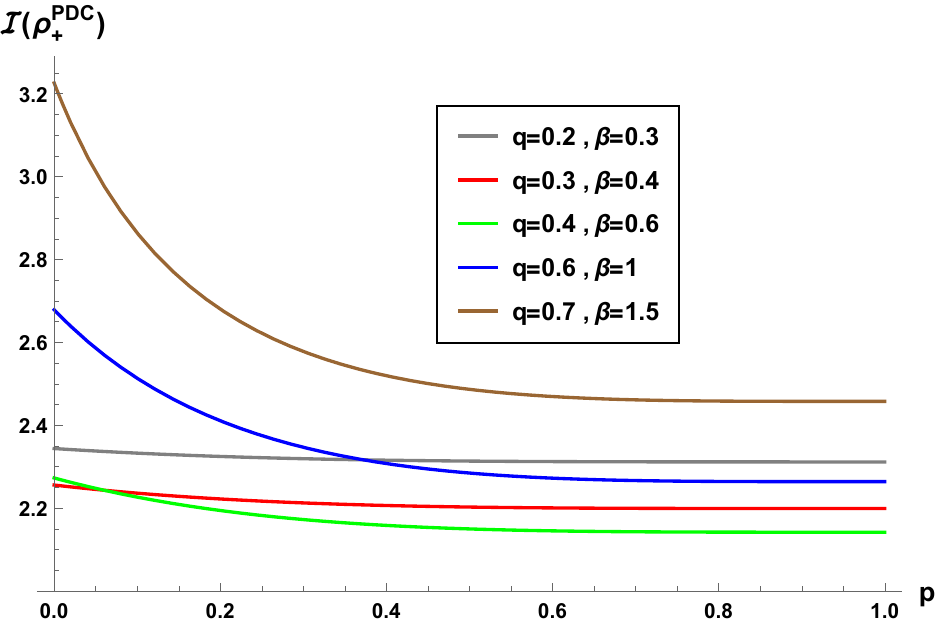}\vfill
				$(c)$
			\end{minipage}}}
	\caption{The evolution of $(a)$ quantum Fisher information, $(b)$ Wootters concurrence, and $(c)$ skew information under phase damping channel versus the decoherence parameter $p$ for the different values of the coherent amplitude $\beta$ with $\alpha=0.5$.}\label{Fig2}
\end{figure}

\subsection{Dynamics of quantum criteria under depolarizing channel}
We now turn our attention to the case of a depolarizing channel effect, applying the Kraus operators via (\ref{DPC}) on the two quasi-Werner states (\ref{+}) and (\ref{-}), it is straightforward to check that the resulting evolved density operator continues to be of $X$-type. Having applied the above results, the quantum Fisher information of the two quasi-Werner states under this noise is written as
\begin{equation}
\mathcal{F}\left(\rho_{\pm}^{\rm DPC} \right)=\frac{s^{4}}{2}\left[\frac{1}{1-qs^{2}}+\frac{2+\left(s^{2}q-1\right)\epsilon^{\pm}}{\left(1+2qs^{2}+q^{2}s^{4}\epsilon^{\pm}\right)}\right],
\end{equation}
with the coefficient $\epsilon^{\pm}$ defined as follows
\begin{equation}\label{}
	\epsilon^{\pm}=1-\frac{n_{\pm}
		^{4}}{N_{\alpha}^{+2}N_{\beta}^{+2}N_{\alpha}^{-2}N_{\beta}^{-2}}-4\kappa_{\pm}^{2}.
\end{equation}
Moreover, one can readily obtain the desired expression of the concurrence entanglement for quasi-Werner states under depolarizing effect as
\begin{equation}
	\mathcal{C}\left(\rho_{\pm}^{\rm DPC} \right)==\max\left\lbrace0,\frac{s^{2}qn_{\pm}^{2}-\left(1-s^{2}q\right)N^{+}_{\alpha} N^{-}_{\alpha} N^{+}_{\beta} N^{-}_{\beta} }{2N^{+}_{\alpha} N^{-}_{\alpha} N^{+}_{\beta} N^{-}_{\beta}}\right\rbrace.
\end{equation}
In a more compact form, the skew information for quasi-Werner states under depolarizing channel is given by
\begin{align}
\mathcal{I}\left(\rho_{\pm}^{\rm DPC} \right)&=\frac{1}{2\left(1-s^{2}q\right)}+\frac{D_{\pm}^{2}s^{4}}{8}\left(E_{\pm}^{-2}-\frac{qs^{2}n_{\pm}^{4}}{2N_{\alpha}^{+2}N_{\beta}^{+2}N_{\alpha}^{-2}N_{\beta}^{-2}}+2qs^{2}\kappa_{\pm}^{2}\right)^{2}\notag\\& +s^{4}\left( 2\kappa_{\pm}^{2}+\frac{n_{\pm}^{4}}{2N_{\alpha}^{+2}N_{\beta}^{+2}N_{\alpha}^{-2}N_{\beta}^{-2}}\right) \left(E_{\pm}-\frac{qs^{2}D_{\pm}}{4}+\frac{q^{2}s^{4}n_{\pm}^{4}F_{\pm}}{8N_{\alpha}^{+2}N_{\beta}^{+2}N_{\alpha}^{-2}N_{\beta}^{-2}}+\frac{q^{2}s^{4}\kappa_{\pm}^{2}F_{\pm}}{2}\right)^{2},
\end{align}
where
\begin{equation}
E_{\pm}=D_{\pm}\varpi_{\pm}=\sqrt{2}\left(1+qs^{2}+\varpi_{\pm}\right)^{-1/2},
\end{equation}
and
\begin{equation}
F_{\pm}=E_{\pm}^{3}\varpi_{\pm}^{-1},\quad \varpi_{\pm}=\sqrt{1+2qs^{2}+q^{2}s^{4}\epsilon^{\pm}}.
\end{equation}

\begin{figure}[h!]
	{{\begin{minipage}[b]{.33\linewidth}
				\centering
				\includegraphics[scale=0.36]{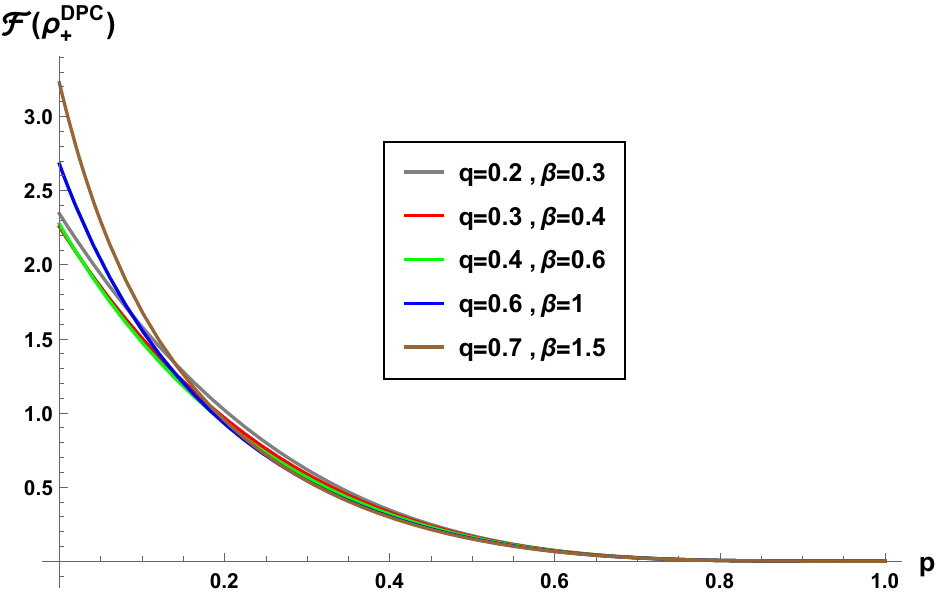}\vfill
				$(a)$
			\end{minipage}\hfill
			\begin{minipage}[b]{.33\linewidth}
				\centering
				\includegraphics[scale=0.36]{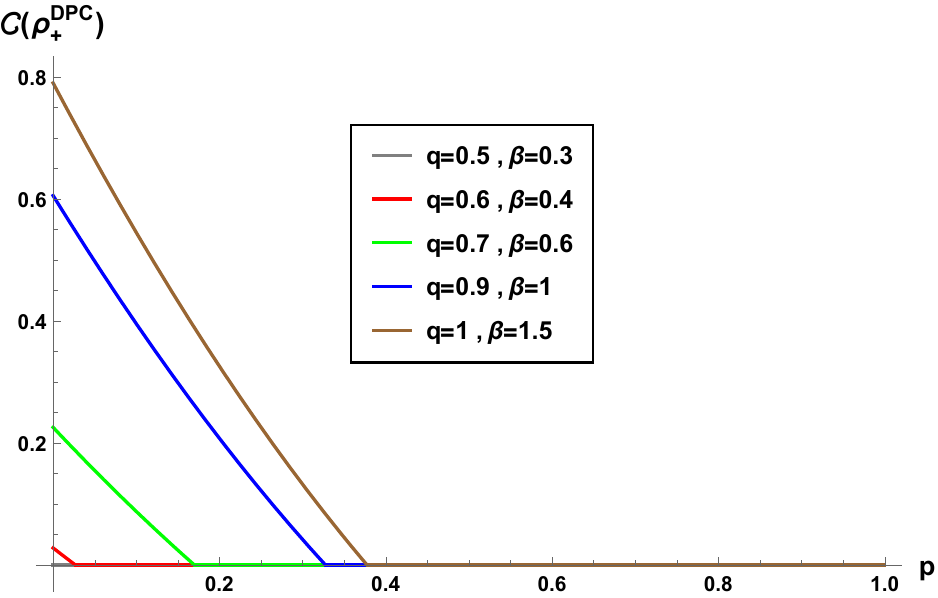}\vfill
				$(b)$
			\end{minipage}\hfill
			\begin{minipage}[b]{.33\linewidth}
				\centering
				\includegraphics[scale=0.36]{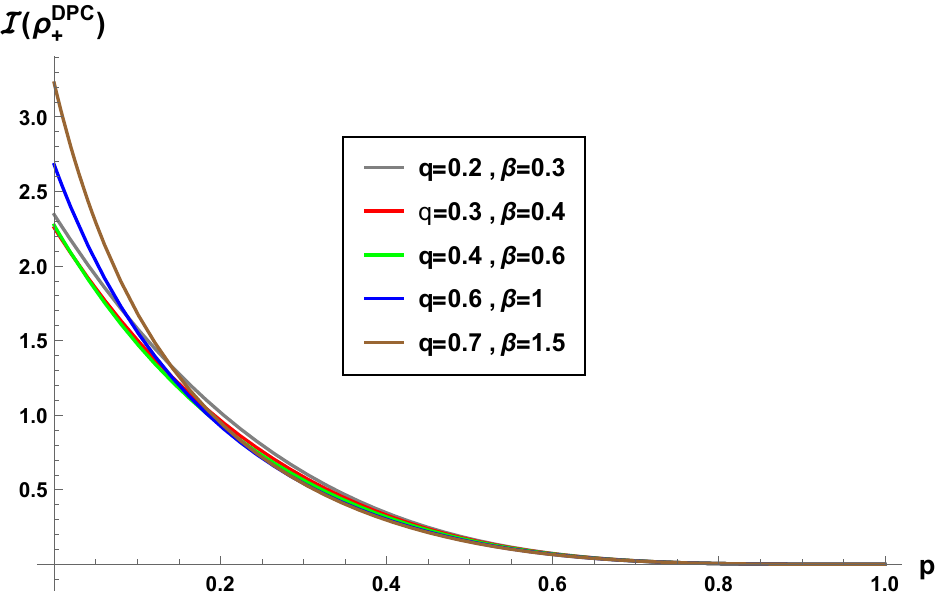}\vfill
				$(c)$
	\end{minipage}}}
	\caption{Dynamics of quantum Fisher information, concurrence and skew information under depolarizing channel with respect to the same parameter sets in Fig.\ref{Fig2}.}\label{Fig3}
\end{figure}
\subsection{Dynamics of quantum criteria under amplitude damping channel}
To implement the Kraus operators of the amplitude damping channel given by (\ref{ADC}) on the two quasi-Werner states, we obtain the analytical expression of the quantum Fisher information under this effect as
\begin{align}
\mathcal{F}\left(\rho_{+}^{\rm ADC}\right)&=\frac{1}{4\mu}
\left[\frac{4\nu^{2}-4(1-s+s^{2}/2)^{2}\xi}{(1-s+s^{2}/2)^{2}+2q\nu+q^{2}\xi}+\left(s^{2}-\frac{s(1-s)n_{+}^{2}\left(N_{\alpha}^{-2}N_{\beta}^{-2}-N_{\alpha}^{+2}N_{\beta}^{+2} \right)}{2N_{\alpha}^{-2}N_{\beta}^{-2}N_{\alpha}^{+2}N_{\beta}^{+2}}\right)^{2} \right]\notag\\&+\frac{s\left(\left(1-s\right)n_{+}^{2}\left(N_{\alpha}^{-2}N_{\beta}^{-2}-N_{\alpha}^{+2}N_{\beta}^{+2} \right)-2sN_{\alpha}^{-2}N_{\beta}^{-2}N_{\alpha}^{+2}N_{\beta}^{+2} \right)^{2}}{4N_{\alpha}^{-2}N_{\beta}^{-2}N_{\alpha}^{+2}N_{\beta}^{+2}\left(N_{\alpha}^{-2}N_{\beta}^{-2}N_{\alpha}^{+2}N_{\beta}^{+2}\left(2-s-2qs\right)+\left(1-s\right)n_{+}^{2}\left(N_{\alpha}^{-2}N_{\beta}^{-2}-N_{\alpha}^{+2}N_{\beta}^{+2} \right)\right)},
\end{align}
for the quasi-Werner states $\rho(\psi^{+}, a)$, where
\begin{equation}
\nu=\frac{\left(2\left(1-s\right)+s^{2}\right)\left(2s^{2} N_{\alpha}^{-2}N_{\beta}^{-2}N_{\alpha}^{+2}N_{\beta}^{+2}-s\left(1-s \right)n_{+}^{2}\left(N_{\alpha}^{-2}N_{\beta}^{-2}-N_{\alpha}^{+2}N_{\beta}^{+2} \right)\right) }{8N_{\alpha}^{-2}N_{\beta}^{-2}N_{\alpha}^{+2}N_{\beta}^{+2}},
\end{equation}
\begin{equation}
\mu=\frac{2q\nu+\left(s^{2}+2\left(1-s\right)\right)^{2}}{2\left(s^{2}+2\left(1-s\right)\right)},\hspace{1cm}{\rm and}\hspace{1cm}\xi=\frac{2\nu}{2\left(1-s\right)+s^{2}}+\frac{s^{2}n_{+}^{2}\left(N_{\alpha}^{+2}N_{\beta}^{+2}-N_{\alpha}^{-2}N_{\beta}^{-2}-4 \right) }{N_{\alpha}^{-2}N_{\beta}^{-2}N_{\alpha}^{+2}N_{\beta}^{+2}}
\end{equation}
and
\begin{equation}
\mathcal{F}\left(\rho_{-}^{\rm ADC}\right)=\frac{2s^{2}\left(1-s\right)+s^{4}\left(1-q\right)}{2\left(1-q\right)\left(s^{2}\left(1-q\right)+4\left(1-s\right)\right)}+\frac{1}{s\left(2-s^{2}\right)+2sq}\left[s^{4}-2\eta+\frac{2\left(s^{3}\left(2-s^{2} \right)+2q\eta\right)^{2}}{s^{2}\left(2-s^{2}+2q \right)+2q^{2}\eta}\right]
\end{equation}
for the quasi-Werner states $\rho(\psi^{-}, a)$, with
\begin{equation}
\eta=\frac{s^{2}}{4}\left[s^{2}-\frac{n_{-}^{4}\left(1+\left(N_{\alpha}^{-2}N_{\beta}^{+2}-N_{\alpha}^{+2}N_{\beta}^{-2} \right)^{2} \right)}{N_{\alpha}^{-2}N_{\beta}^{-2}N_{\alpha}^{+2}N_{\beta}^{+2}\left(1+4N_{\alpha}^{-2}N_{\beta}^{-2}N_{\alpha}^{+2}N_{\beta}^{+2} \right)} \right].
\end{equation}
Herein, the concurrence entanglement in the evolved quasi-Werner state $\rho_{+}^{\rm ADC}$ is
\begin{equation}
	\mathcal{C}\left(\rho_{+}^{\rm ADC}\right)=\max\left\lbrace0,\frac{sqn_{+}^{2}}{2N^{+}_{\alpha} N^{-}_{\alpha}N^{+}_{\beta} N^{-}_{\beta}}-s\left(1+\left(1-s\right)q\kappa_{+}\right)+\frac{s^{2}}{2}\left(1+q\right)\right\rbrace,
\end{equation}
and for the evolved quasi-Werner state $\rho_{+}^{\rm ADC}$, it is given by the following expression
\begin{equation}
	\mathcal{C}\left(\rho_{+}^{\rm ADC}\right)=\max\left\lbrace0,\frac{sqn_{-}^{2}}{2N^{+}_{\alpha} N^{-}_{\alpha} N^{+}_{\beta} N^{-}_{\beta}}-\frac{1}{2}\sqrt{s^{2}\left(1-q\right)\left(4\left( 1-s\right)+s^{2}\left(1-q\right) \right)}\right\rbrace.
\end{equation}
Afterward, the analytical expression of the skew information for the quasi-Werner state $\rho(\psi^{+}, q)$ under amplitude damping channel can be determined from Eq.(\ref{52}) as
 \begin{align}
&\mathcal{I}\left(\rho_{+}^{\rm ADC}\right)=\frac{{\cal A}^{2}s^{2}}{8}\left(s{\cal B}^{-2}+{\cal X}-2qs\kappa_{+}^{2}-\frac{qsn_{+}^{4}}{2N_{\alpha}^{+2}N_{\beta}^{+2}N_{\alpha}^{-2}N_{\beta}^{-2}}\right)^{2}+\frac{s\left(2\left(1-s \right)\kappa_{+}-s\right)^{2}}{2\left(2-s\right)+2q\left(2\left(1-s\right)\kappa_{+}-s\right)}+\notag\\& \frac{s^{2}n_{+}^{4}}{2N_{\alpha}^{+2}N_{\beta}^{+2}N_{\alpha}^{-2}N_{\beta}^{-2}}\left({\cal B}-\frac{qs{\cal A}}{4}\left(s-2\left(1-s \right)\kappa_{+}\right)+\frac{q{\cal B}^{2}{\cal A}}{2}{\cal R}\right)^{2}+\frac{1}{8}\left(-s{\cal A}{\cal Y}+sn_{+}^{2}{\cal B}\left(2+q{\cal B}{\cal A}{\cal R}\right)-2\left(1-s\right){\cal B}^{2}{\cal A}{\cal R}\right)^{2},
 \end{align}
 where
\begin{equation}
\Sigma=\left(s^{2}-2s\left(1-s\right)\kappa_{+}\right)^{2}-s^{2}\kappa_{+}^{2}-\frac{s^{2}n_{+}^{4}}{N_{\alpha}^{-2}N_{\beta}^{-2}N_{\alpha}^{+2}N_{\beta}^{+2}},\hspace{1cm}{\cal R}=\frac{qs^{2}n_{+}^{2}}{4N_{\alpha}^{-2}N_{\beta}^{-2}N_{\alpha}^{+2}N_{\beta}^{+2}}+\kappa_{+}\left(qs^{2}\kappa_{+}+s\left(s-1\right)\right),
\end{equation}
\begin{equation}
\Gamma=\frac{1}{4}\left(s^{2}-2s+2\right)\left(s^{2}\left(2q+1 \right)-2s +2\right)+\left(2s-s^{2}-1\right)\left(qs\left(1-s \right)\kappa_{+}+1\right),  
\end{equation}
\begin{equation}
{\cal A}=2\left(4\Gamma+q^{2}\Sigma\right)^{-\frac{1}{2}}\left(2\left(1-s \right)+s^{2}+q\left(s^{2}-2\left(1-s\right)\kappa_{+}\right)+\sqrt{4\Gamma+q^{2}\Sigma}\right)^{\frac{1}{2}},\hspace{0.75cm}{\cal B}=\frac{{\cal A}}{2}\sqrt{4\Gamma+q^{2}\Sigma},  
\end{equation}
\begin{equation}
{\cal X}=2\kappa_{+}\left[\left(1-s\right)\left(1-\frac{1}{{\cal B}^{2}}\right)-qsn_{+}^{2}\kappa_{+}\right],\hspace{1cm}{\cal Y}=\left(s\left(1+q\kappa_{+} \right)-1\right)\left(s-2\left(1-s\right)\kappa_{+}\right),   
\end{equation}
and for the quasi-Werner state $\rho(\psi^{-}, q)$ as
\begin{align}
\mathcal{I}\left(\rho_{-}^{\rm ADC}\right)&=\frac{\Theta^{2}s^{4}}{8}\left(\Upsilon^{-2}-\frac{qn_{-}^{4}}{2N_{\alpha}^{+2}N_{\beta}^{+2}N_{\alpha}^{-2}N_{\beta}^{-2}}-2q\kappa_{-}^{2}\right)^{2}+\frac{\varsigma^{2} s^{4}}{8}\left(\zeta^{-4}+(s-1)^{2}\right)+\notag\\&\left(\frac{s^{2}n_{-}^{4}}{2N_{\alpha}^{+2}N_{\beta}^{+2}N_{\alpha}^{-2}N_{\beta}^{-2}}+2s^{2}\kappa_{-}^{2}\right)\left(\Upsilon-\frac{qs^{2}\Theta}{4}+\frac{q^{2}s^{2}\Upsilon^{2}\Theta}{2}\left(\kappa_{-}^{2}+\frac{n_{-}^{4}}{4N_{\alpha}^{+2}N_{\beta}^{+2}N_{\alpha}^{-2}N_{\beta}^{-2}}\right)\right)^{2},  
\end{align}
where the quantities $\zeta$, $\varsigma$, ${\cal D}$, $\Upsilon$ and $\Theta$ are given by
\begin{equation}
\zeta=\sqrt{2}\left(2\left(1-s\right)+s^{2}\left(1-q\right)+s\sqrt{\left(1-q\right)\left(4\left(1-s\right)+s^{2}\left(1-q\right)\right)}  \right)^{-\frac{1}{2}},\hspace{0.5cm}\varsigma=\frac{s\zeta}{2}\sqrt{\left(1-q\right)\left(1-s+s^{2}\left(1-q\right)\right)},
\end{equation}
\begin{equation}
{\cal D}=s^{2}-\frac{n_{-}^{4}}{N_{\alpha}^{+2}N_{\beta}^{+2}N_{\alpha}^{-2}N_{\beta}^{-2}}-4\kappa_{-}^{2},\hspace{0.5cm}\Upsilon=\sqrt{2}\left(s\left(2+s\right)-qs^{2}+\sqrt{s^{2}\left(2-s \right)\left(2qs+2-s\right)+q^{2}s^{2}{\cal D}}\right)^{-\frac{1}{2}}, 
\end{equation}
\begin{equation}
\Theta=\Upsilon\sqrt{2}\sqrt{s^{2}\left(2-s \right)\left(2qs+2-s\right)+q^{2}s^{2}{\cal D}}.
\end{equation}

\begin{figure}[h!]
	{{\begin{minipage}[b]{.33\linewidth}
				\centering
				\includegraphics[scale=0.36]{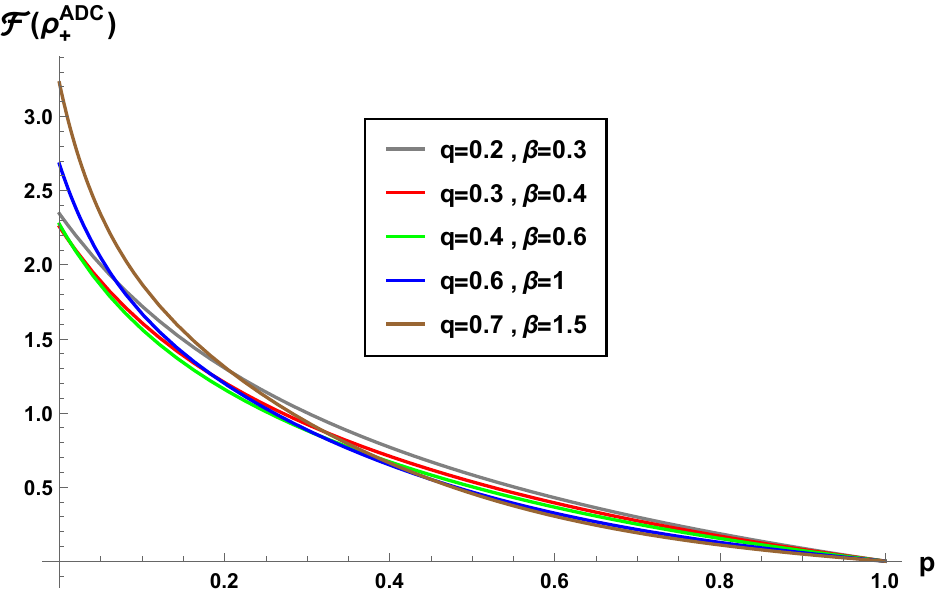}\vfill
				$(a)$
			\end{minipage}\hfill
			\begin{minipage}[b]{.33\linewidth}
				\centering
				\includegraphics[scale=0.36]{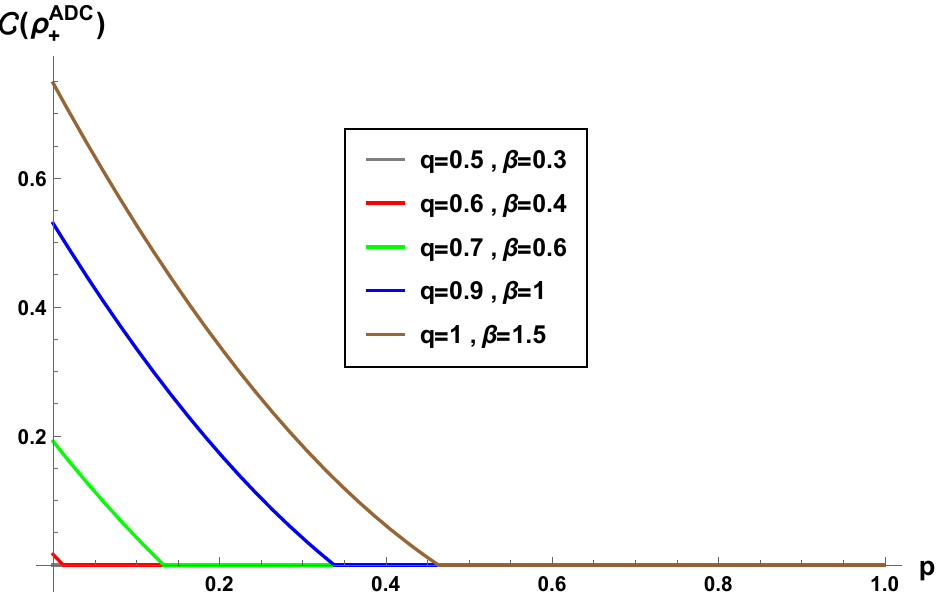}\vfill
				$(b)$
			\end{minipage}\hfill
			\begin{minipage}[b]{.33\linewidth}
				\centering
				\includegraphics[scale=0.36]{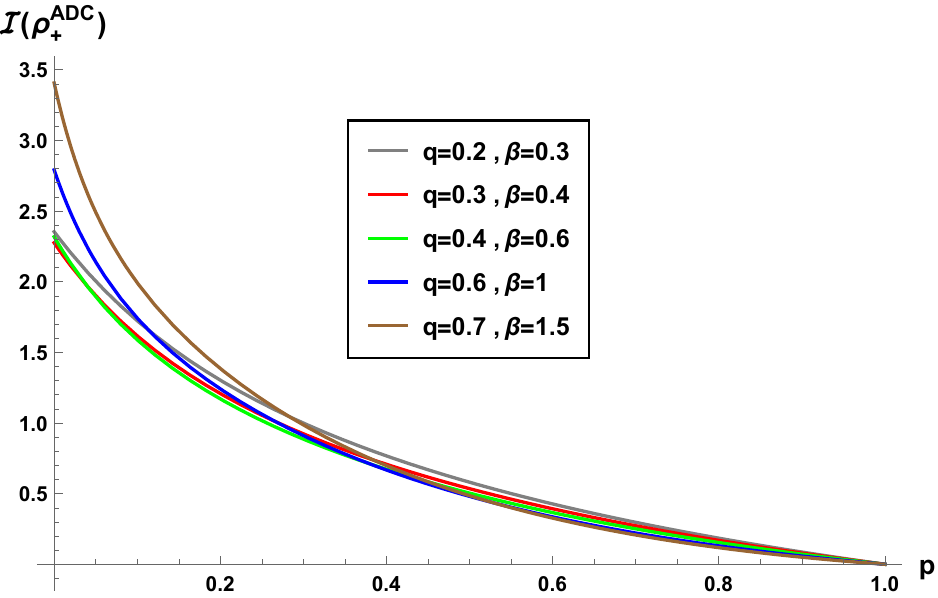}\vfill
				$(c)$
	\end{minipage}}}
	\caption{Plot of the quantum criteria versus the amplitude damping parameter $p$ for the same parameters as in Fig.\ref{Fig2}}\label{Fig4}
\end{figure}
In order to demonstrate the decoherence effect on the different kinds of quantum physical quantities, we display in Fig.\ref{Fig2}, Fig.\ref{Fig3} and Fig.\ref{Fig4} the dynamics behaviors of the quantum Fisher information (panel ($a$)), concurrence (panel ($b$)) and Wigner-Yanase skew information (panel ($c$)) with respect to the decoherence parameter $p$ for the markovian phase-damping, depolarizing and amplitude-damping channels, respectively. We can see on the plots, and for all the channels considered, that the coherent amplitude and the mixing parameter considerably influence the dynamic behavior of these quantum criteria during the temporal evolution. Furthermore, these figures reveal that the maximum of all quantum criteria are obtained for the largest values of both coherent amplitude and mixing parameter, and that they are affected by the mixedness induced by small values of mixing parameter $q$. These results indicate that controlling these parameters under decoherence effect greatly benefits the physical quantities of the quasi-Werner states. We also find that those quantities decrease with increasing decoherence parameter $p$, which has a negative role for their improvement.\par

Furthermore, a comparison of the evolution of these measures reveals a strong decrease in concurrence compared to the quantum Fisher and skew information. This suggests that the concurrence dynamics under the decoherence process has a strong destruction while the Fisher and skew information show a gradual decrease. We further note that these quantities are more robust in the phase damping channel compared to the other quantum channels. Moreover, quantum Fisher and skew information exhibit similar behavior for the three different decoherence channels. On the other hand, the Fisher and skew information are maximal when the entanglement is maximal, and in this value, the quasi-Werner states offer the best estimation efficiency. This shows that skew information fulfills the same role as quantum Fisher information in quantum estimation theory.

\section{Concluding Remarks}
To summarize, the skew information as a measure of the theoretical information content in a given quantum state, and the quantum Fisher information defined via symmetric logarithmic derivatives, have some common properties, both playing an important role in quantum information theory and establishing the interpretation in quantum uncertainty. They are both useful in the quantification of quantum resources such as quantum coherence \cite{Yu2017,Li2021}, asymmetry \cite{Sun2021,Li2020} and quantum correlation \cite{SlaouiD2019,Dhar2015,Girolami2014}. Moreover, understanding these informational quantifiers in quantum systems has practical significance in quantum estimation strategies. However, the interplay between these two concepts appears to have attracted little attention.\par

Our main result provides a complete characterization of such notions and gives a detailed analytical derivation for an arbitrary two-qubit $X$-state. As an illustration, we applied these derived analytical expressions in the two quasi-Werner states composed of two bipartite superposed coherent states. As well, we investigated their dynamics when exposed to various decoherence effects via phase damping, amplitude damping and depolarization channels. We have proved that quantum Fisher information and skew information perform the same behavior under decoherence channels and exhibit somewhat similar properties. Further, the maximum skew information coincides with the maximum amount of entanglement in the two quasi-Werner states, which is quantified via the Wootters concurrence. These results are in agreement with the results of the quantum Fisher information and show the role of quantum entanglement to improve the efficiency and accuracy of quantum metrology protocols.\par

{\bf Acknowledgment:} The authors are very grateful to the referees for their constructive remarks and for their insightful comments and observations that improved the quality of this work paper.

\end{document}